\documentclass[aps,prd,twocolumn,groupedaddress,showpacs,amsmath,amssymb]{revtex4-1}

\usepackage{graphicx}
\usepackage{bm,color}

\newcommand{\be}{\begin{eqnarray}}
\newcommand{\ee}{\end{eqnarray}}

\def\e{\epsilon}

\def\D{\Delta}

\begin{document}

\title{Topological order versus many-body localization in periodically modulated
spin chains 
}
\author{Takahiro Orito$^1$}
\author{Yoshihito Kuno$^2$}
\author{Ikuo Ichinose$^1$}
\affiliation{$^1$Department of Applied Physics, Nagoya Institute of Technology, Nagoya, 466-8555, Japan}
\affiliation{$^2$Department of Physics, Graduate School of Science, Kyoto University, 
Kyoto, 606-8502, Japan}

\date{\today}

\begin{abstract}
In this paper, we study periodically modulated $s=1/2$ spin chain in a linear gradient potential (LP)
that is generated by an external magnetic field.
In the absence of the LP, the system has topological states that exhibit a magnetization plateau
for a uniform external magnetic field.
These topological states have a finite integer Chern number and their stability is clarified by an equivalent
spinless fermion system derived by a Jordan-Wigner transformation.
We show that the LP, which is nothing but a constant electric field in the spinless fermion system, 
destabilizes the topological states, because it induces localization called Wannier-Stark (WS) localization.
We clarify the phase diagram in the presence of the LP and on-site diagonal disorder.
To this end, we carefully study edge excitations under the open boundary condition,
which are a hallmark of the topological order.
We find a very interesting phenomenon indicating existence of a quasi-edge modes 
that take the place of the genuine edge modes in certain parameter regions.
This is a precursor of the WS localization realized in topological states.
Finally, we investigate many-body localization induced by a sufficiently strong LP or disorder. 
To this end, we study the energy-level statistics for whole energy levels, and find unexpected
extended-state regimes located in intermediate potential-gradient and weak on-site disorder
regimes.
We verify this phenomenon by calculating variance of the entanglement entropy.
The present system is closely related to quantum Hall state in two dimensions, and therefore
our findings can be observed not only in experiments on ultra-cold atomic gases but also quantum
Hall physics.
\end{abstract}
\maketitle


\section{Introduction}

Both topological order and many-body localization (MBL) are one of the most
important topics in condensed matter phase these days.
Recent experiments on cold atoms succeeded in
observing fundamental signals of the conventional topological phase \cite{Jotzu, Asteria} 
and also ergodicity breaking dynamics of MBL~\cite{Schreiber,Choi}. 
These experimental developments stimulated theoretical study of constructing fundamental models 
for both topological phase~\cite{Cooper,Ozawa} and MBL~\cite{Abanin}.
The experimental success also enhances interest on the interplay of the topological phase and MBL. 
Recently, interesting theoretical works on this subject were given in
Refs.~\cite{Huse0,Chandran,Bauer,Parameswaran}. 
In the models studied there, the topological phase, especially edge modes, are to be protected 
from localization by disorders, and even highly excited states are expected to possess 
the nature of the topological state.
Soon after these works, a few numerical studies~\cite{Bahri,Decker,Kuno} have verified the conjecture. 
As another interesting work, MBL induced not by disorder but by a linear gradient potential (LP) 
was recently discovered~\cite{Schulz,Refael}. 
This MBL phenomenon without disorders comes from the Wannier-Stark (WS) 
localization of the single-particle system~\cite{Kolovsky}.
As the system under the LP possesses the translational symmetry, 
its entanglement properties are to be different from the conventional MBL systems~\cite{Schulz}. 
As the LP is easier to be produced in experiments on cold atomic gases than random potentials,
cold atom systems under the LP provide a good playground for study on the interplay of topological
order and localization.

In this paper, we investigate the relationship between a conventional 
one-dimensional topological model 
characterized by the Chern number~\cite{Thouless} and the WS localization mentioned above.
To this end, we consider a periodically modulated $s=1/2$ spin chain~\cite{Hu,Chen}. 
This model has magnetization plateaus~\cite{Oshikawa}, corresponding to a topological phase. 
Through dimensional extension~\cite{QHS}, the topological phase is characterized
as a Chern insulator in the extended two-dimensional space, 
where the Chern number takes an integer values~\cite{Hu}. 
We consider to apply the LP to the model and study the effect of the WS localization
in the topological phase with a magnetization plateau. 
Here, we would like to emphasize that the topological phase is usually related with a property of 
the ground state wave function of the system.
On the other hand, the localization including MBL is properties of whole energy eigenstates, i.e., 
all energy eigenstates of the system are related with localization of the system.
From this point of view, we will investigate both the ground state topological properties characterized 
by the Chern number and edge excitations, and also many-body eigenstates in the whole energy
spectrum characterizing localization of the system. 
In particular, we study robustness of the ground-state topological properties against the LP, 
and how the WS localization influences the topological ground state. 
In this paper, we mainly use a numerical exact diagonalization (ED)~\cite{ED1,ED2,ED3,ED4}, 
since we need to investigate properties of the whole energy eigenstates. 
Numerically, we will clarify the phase diagram in the presence of the LP and on-site disorder, 
by the energy-level statistics~\cite{Alet,Janarek} to detect the WS localization, 
and investigate in detail the behavior of the topological edge modes under the LP.

This paper is organized as follows.
In Sec.~II, we introduce the target model and summarize the previous works that are relevant
to the present work.
In particular, we explain topological phase and the WS localization induced by the LP.
After the summary of the previous works, we explain the purpose of the present study.
In Sec.~III, we study the topological phase in the presence of the LP and on-site disorder.
We first show that there exists a critical gradient for destruction 
of the ground state topological phase by calculating the Chern number.
Then, applying the open boundary condition,
we investigate the edge excitations in detail, and show that a precursory phenomenon of 
the WS localization appears in the behavior of the edge excitations.
In Sec.~IV, MBL is studied by the energy-level statistics to obtain the phase boundary
of the ergodic and MBL states.
In the weak-disorder regime, phase diagram exhibits an unexpected structure,
which comes from the interplay between the LP and the modulated exchange coupling.
We calculate entanglement entropy to investigate this structure in detail, and discuss its
possible physical picture.
Section V is devoted for conclusion and discussion.

\section{model and summary of previous works}\label{model}

In this work, we consider the following $s=1/2$ spin chains with a periodically
modulated exchange coupling, 
\be
H_{\rm PM}=\sum_i J_i (\vec{S}_i\cdot \vec{S}_{i+1})-h\sum_iS^z_i,
\label{HPM}
\ee
with
\be
J_i=[1-\lambda \cos (2\pi\alpha i+\delta)],
\label{Ji}
\ee
where $\vec{S}_i=(S^x_i,S^y_i,S^z_i)$ is spin operator at site $i$, $\alpha$ is 
a rational number and $\delta$ takes an arbitrary real number.
In the following, we mostly take $\lambda=0.8$, whereas an external magnetic field
$h$ is adjusted to realize a constant magnetization $m_z=\langle S^z_i\rangle/L$,
where $L$ is the system size. 
Previous works revealed that the model $H_{\rm PM}$ [Eq.~(\ref{HPM})] 
has topological states with a magnetization plateau~\cite{Oshikawa,Hu,Lado}.
These states have a non-vanishing Chern number, which is defined by the 
two-dimensional space $(\delta, \theta)$ with the boundary condition on 
wave functions such as $\psi_{L+1}=e^{i\theta}\psi_1$. 
$\theta$ is a twist phase. 
The origin of the topological phase will be explained in Sec. III B.
We have verified the existence of various topological phases by varying $L, \alpha$ and 
$h$~\cite{OHKI}. 
In what follows, we shall concentrate on $\alpha=1/3$ 
since for non-interacting fermion picture, the system exhibits topological three bulk band.

On the other hand, MBL of spin chains with a {\em uniform
exchange coupling} ($J$) and in the presence of 
a linear gradient potential as well as a random magnetic field ($\{h_i\}$) was investigated \cite{Schulz,Refael},
Hamiltonian of which is given by,
\be
&&H_{\rm WS}=H_{\rm PM}|_{J_i=J}+\Delta H,  \label{HStark} \\
&& \Delta H=\sum_i[FiS^z_i+h_iS^z_i],
\ee
where $F$ is the gradient of the linear potential.
It is known that a $XY$ counterpart of the uniform $H_{\rm WS}$  
reduces to a free fermion model in a uniform electric field of strength $F$
by a Jordan-Wigner (JW) transformation, and
{\em single-body states are all localized in the vicinity of each site $i$} for small disorders $\{ h_i \}$
and $F\neq 0$ as the LP dominates the hopping energy~\cite{Refael}.
This phenomenon is called WS localization.
Anti-ferromagnetic (AF) coupling $\sum_i S^z_i\cdot S^z_{i+1}$ in $H_{\rm PM}|_{J_i=J}$
induces a repulsion between the JW spinless fermions, and it can change 
the WS-localized states to ergodic ones for weak $F$ in the many-body system 
and the localization transition point is shifted to larger value of $F$ compared to the single-body system.
In Ref.~\cite{Refael}, the model $H_{\rm WS}$ was studied for disorder
$h_i\in [-W,W]$, and a phase diagram in the $(F-W)$ plane was obtained.
In a regime of small $F$ and $W$, ergodic states form, whereas MBL states appear
for sufficiently large $F$ and/or $W$. 

In this work, we shall focus on the model defined by the following Hamiltonian,
\be
H_{\cal T}=H_{\rm PM}+\Delta H, 
\label{HT}
\ee
with the boundary condition such as $\psi_{L+1}=e^{i\theta}\psi_1$.
For most of cases, we put $\theta=0$, whereas $\theta$ is varied as $\theta\in[-\pi,\pi]$
in the study of topological properties such as the Chern number. 
[Instead of imposing the twist boundary condition on the wave function,
we change the term in the Hamiltonian of
Eq.~(\ref{HPM}) as ${S}^+_{L}{S}_{1}^-+\mbox{h.c.}\to {S}^+_{L}{S}_{1}^-e^{i\theta}+\mbox{h.c.}$
with the periodic boundary condition, where $S^{\pm}_j=S^x_j \pm i S^y_j$~\cite{twist}.]
This boundary condition causes no problems as long as we consider
a finite system, whereas for $L\to \infty$, $LF\to \infty$ for a nonvanishing $F$,
and an infinite deference in the potential energy appears between two edges.
We can change the linear gradient potential to a $V$-shape one, which is given
by $\Delta H_V=\sum_i[F|i-{L \over 2}|S^z_i+h_iS^z_i]$.
However, we have verified that the $V$-shape potential produces similar results
as the linear gradient potential at least for systems with moderate system sizes.
Later, we shall show certain quantities calculated in the model
$H_{\cal TV}=H_{\rm PM}+\Delta H_V$
to verify that the linear gradient and $V$-shape potentials produce essentially the same results. 

Purpose of the present work is three-fold; \\
(i) We investigate how the ground state evolves from the topological state as
$F$ and/or $W$ increase, and how the topological order and localization interplay with each other.
(ii) If a finite regime of the topologically ordered state exists for finite $F$'s, 
we study how the state is affected by the gradient potential. 
In other words, ``precursory localization phenomenon" exists or not in the topological phase.
(iii) We obtain a phase boundary of MBL state that forms as $F/W$ increase and 
compare the resultant MBL state with that in the uniform exchange-coupling system. 
We find that the phase diagram of the present model exhibits unusual structure,
revival of extended states in intermediate-$F$ regimes.
In order to verify this observation, we employ variance of the entanglement entropy
as an `order parameter' of MBL.

Obviously, the first and second subjects in the above are mainly related with properties 
of the ground state and low-energy states of the system.
On the other hand, for the subject (iii), we consider the whole energy
spectrum by introducing the normalized energy 
$\epsilon\equiv (E-E_{\rm Min})/(E_{\rm Max}-E_{\rm Min})$,
where $E_{\rm Max}(E_{\rm Min})$ is the maximum (minimum) energy eigenvalue for
fixed values of $F$ and $W$.
For each problem in the above, we shall clarify spatial configurations of spins
for typical states, and show that the description in terms of the JW fermion 
is sometimes useful to understand the results obtained by the numerical methods.


\section{Topological phase in gradient potentials and disorder}\label{topology}

\subsection{Phase diagram}\label{PD1}

\begin{figure}[t]
\centering
\begin{center}
\includegraphics[width=6cm]{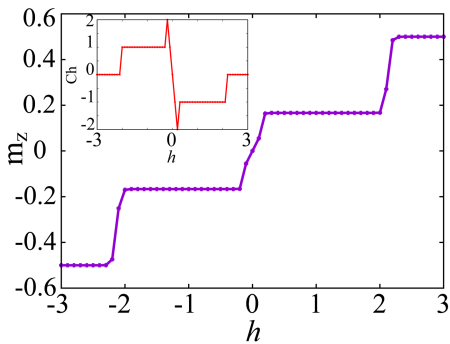}
\end{center}
\caption{Magnetization $m_z$ as a function of uniform external field $h$.
There exist plateaus, and they have a definite Chern number as shown in inset. 
$\alpha={1\over 3}$, $\lambda=0.8$ and system size $L=18$.
We focus on the state of the plateau with $m_z={1\over 6}$ in the present work.
}
\label{magneticPl}
\end{figure}

\begin{figure}[t]
\centering
\begin{center}
\includegraphics[width=6cm]{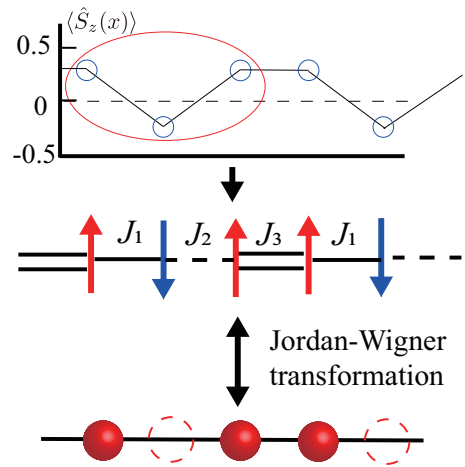}
\end{center}
\caption{Schematic picture of spin configuration and corresponding JW fermion.
${N\over L}={2\over 3}, \ \alpha={1 \over 3}$ and $m_z={1 \over 6}$ case. 
Real fermion states are given by superposing similar states to the displayed one 
as a result of quantum fluctuations. 
}
\label{JWfermion1}
\end{figure}

In this section, we study the topological states in the spin model, $H_{\cal T}$, 
introduced in Sec.~\ref{model} [Eq.~(\ref{HT})].
In order to search topological states, we first calculated the magnetization, 
$m_z={1 \over L}\sum_i\langle S^z_i\rangle$, as a function of $h$ for $F=W=0$ 
by the Lanczos algorithm of the ED~\cite{ED5}.
We show the result for $\alpha=1/q$ with $q=3$ in Fig.~\ref{magneticPl}, which clearly exhibits 
the location of plateaus corresponding to topological states. 
We numerically verified that the magnetization $m_z$ is independent of the values 
of $\delta$ and $\theta$, i.e., locations of topological states are the same for all $\delta$'s and $\theta$'s.
For $\alpha=1/q$, the unit cell is composed of $q$ sites of the original lattice.
All states in the plateaus have clear schematic picture in terms of the JW fermion,
e.g., the state of $m_z={1 \over 6}$ for $L=18$ corresponds to configurations in which exactly two
JW fermions reside in each unit cell as $S^z_i=n_i-{1\over 2}$ 
where $n_i$ is the number operator of the JW fermion $c_i$, $n_i=c^\dagger_i c_i$.
See Fig.~\ref{JWfermion1}.
Locations of fermions fluctuate in the cell due to the quantum effect, and as a result, an
extended many-body state forms.

By applying the linear gradient potential to the system, we expect that a MBL 
state forms for $F>F_{c2}$ where $F_{c2}$ is critical gradient for the WS localization
in the present many-body system.
In order to see destruction of the topological ground state, on the other hand,
we calculate the Chern number 
as a function of $F$ by using the method of the discretized Chern number~\cite{Fukui}, 
and denote the critical gradient of the linear
potential with $F_{c1}$ at which the Chern number tends to vanish.
In general, $F_{c1}$ is different from $F_{c2}$.
It is one of the purposes of the preset work to estimate $F_{c1}$ and $F_{c2}$.

\begin{figure}[t]
\centering
\begin{center}
\includegraphics[width=6cm]{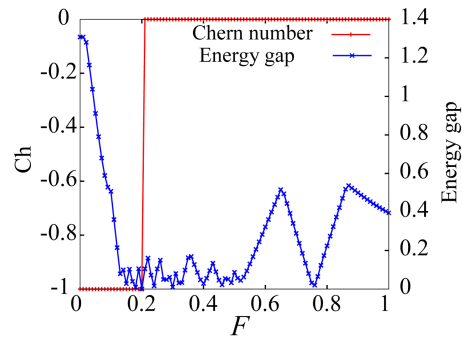}
\end{center}
\vspace{-0.5cm}
\caption{Chern number as a function of $F$ for the case of without
disorder ($W=0$).
At $F=0.2$, the Chern number sharply changes its value indicating $F_{c1}(W=0)=0.2$.
Energy gap between the ground state and the first-excited state
is also shown.
The energy gap collapses between $F=0.2$ and $0.45$.
The system size $L=18$
}
\label{CHN1}
\end{figure}

We show calculations of the Chern number for 
$m_z={1 \over 6}$~\cite{ED6} and $W=0$ as a function of $F$ in Fig.~\ref{CHN1}.
System-size dependence is also examined in appendix A.
From the numerical calculations, we obtain estimation such as $F_{c1}\simeq 0.2$ for $W=0$.
Energy gap between the ground state and the first-excited state is also
shown in Fig.~\ref{CHN1}.
The calculation shows that the energy gap collapses from $F\simeq 0.2$ to $\simeq 0.45$
exhibiting an oscillating behavior.
Under sufficiently strong gradient potential, dynamics of spin tends to get frozen.
We expect that the states between $F\simeq 0.2$ and $\simeq 0.45$ exist in a critical region. 
One may think that $F_{c2}(W=0)\simeq 0.45$ and MBL states exist for $F\gtrsim 0.45$ 
from the above observation.
On the other hand, collapse of energy gap usually indicates localization in the case of random disorders.
The problem of localization will be studied in detail in Sec.~\ref{MBL}
by calculating the energy-level statistics, dipole moment, etc.

\begin{figure}[t]
\centering
\begin{center}
\includegraphics[width=6cm]{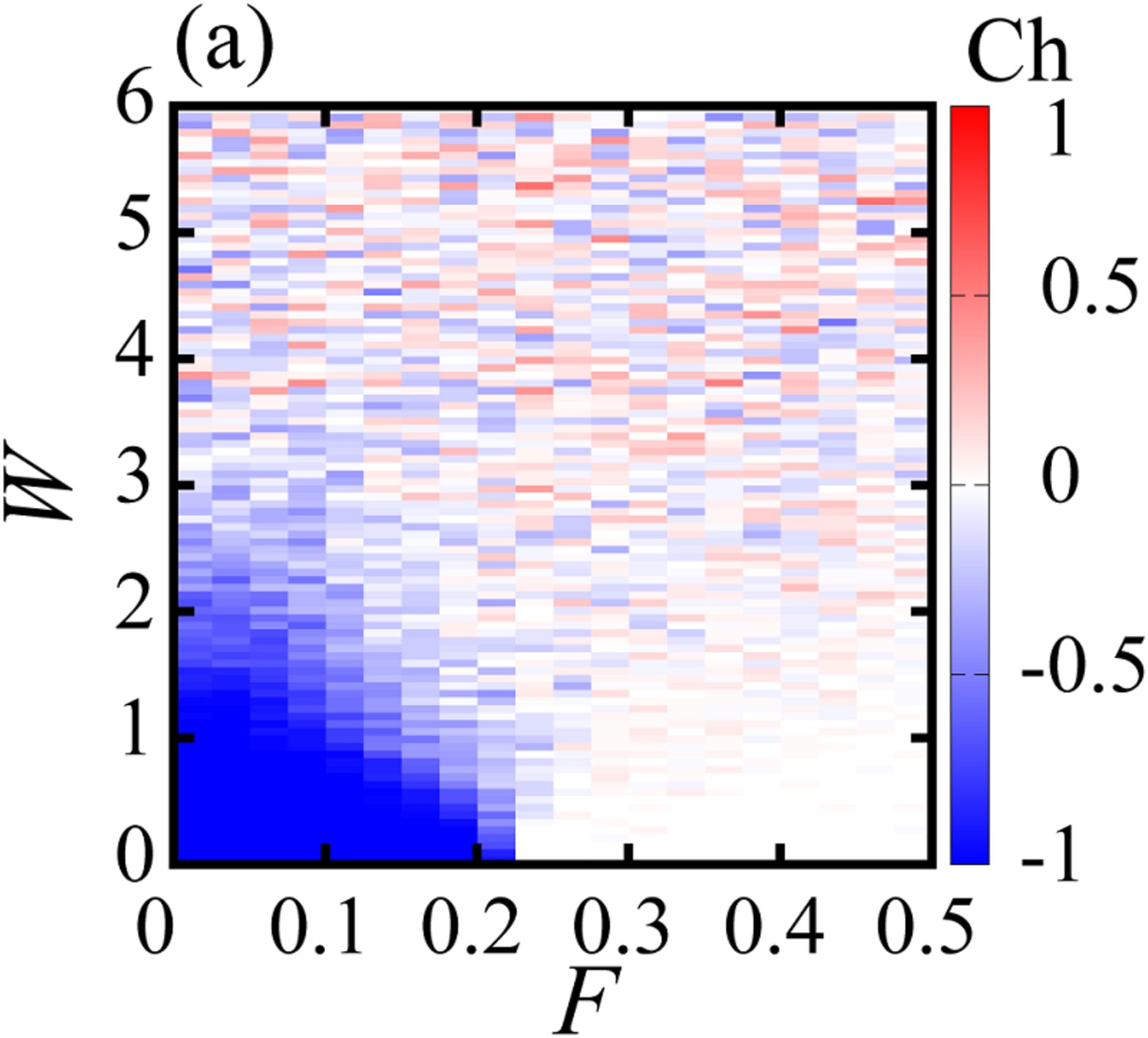}
\includegraphics[width=6cm]{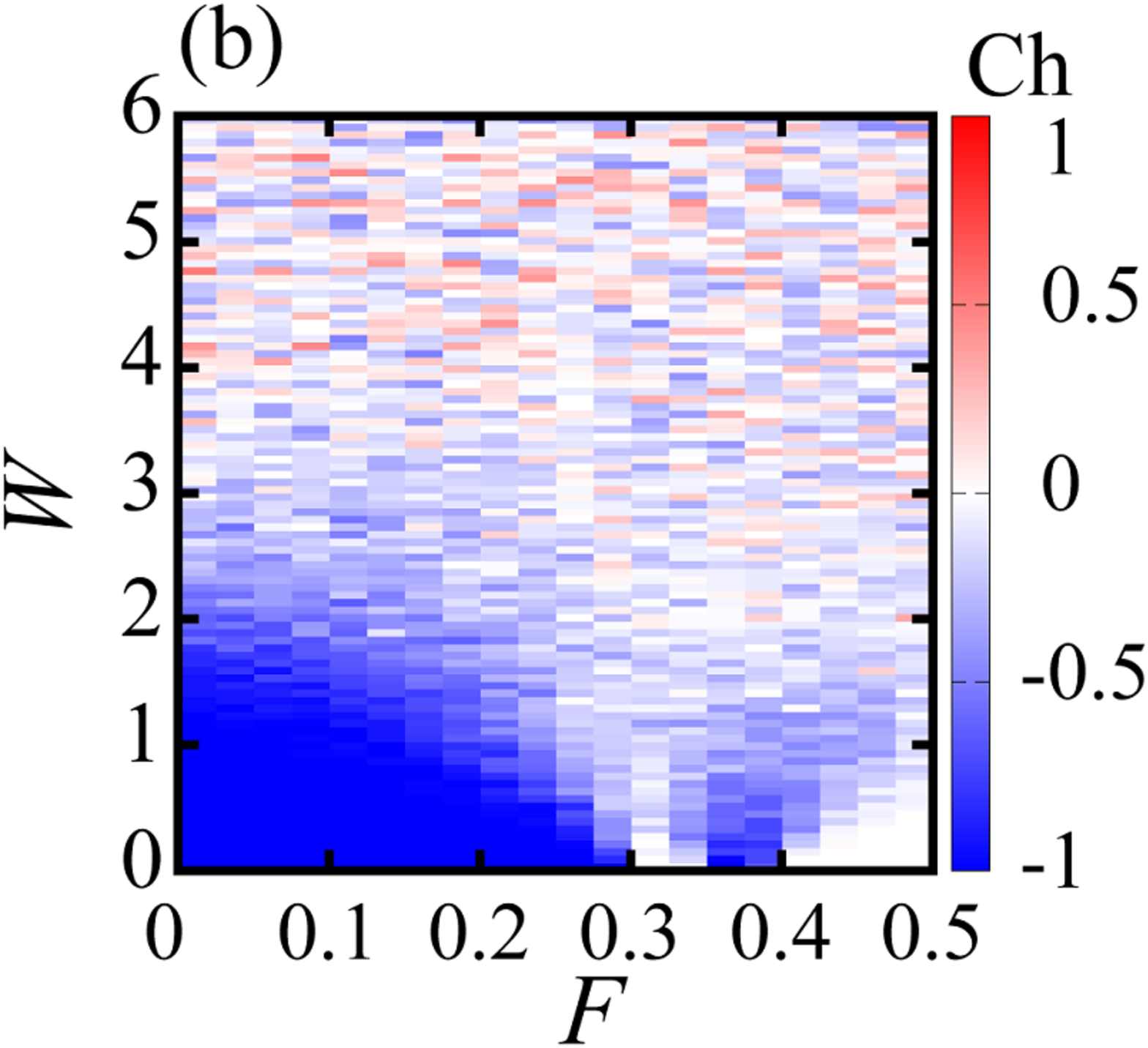}
\end{center}
\vspace{-0.5cm}
\caption{Phase diagram of the topological phase obtained by Chern number. 
(a) Phase diagram for the linear $F$-potential.
(b) Phase diagram for the $V$-shape $F$-potential.
System size $L=18$ and 50 realizations of disorder.
}
\label{WFphase1}
\end{figure}

We studied the system with various values of $W$
by the ED, and obtain $F_{c1}$ as a function of $W$, $F_{c1}(W)$.
For the case of a finite disorder $W$, value of the Chern number, Ch$(W)$, depends on
realizations of disorder $\{h_i \}$, i.e., Ch$(W)=0$ for certain $\{h_i \}$, whereas
Ch$(W)$= a non-vanishing integer for another $\{h_i \}$ with the same $W$.
In Fig.~\ref{WFphase1}, we exhibit the ground state phase diagrams of the topological phase 
in the $(F-W)$-plane obtained by using 50 realizations of  $\{h_i \}$ for each $W$.
In the critical regime in the $(F-W)$ plane, the averaged value of Ch$(W)$ is fractional.
This means that states with Ch$(W)=-1$ and those with Ch$(W)=0$ both
appear as a result of random disorder samples.
Among them, the ground states with Ch$(W)=-1$ have the topological phase nature,
whereas those with Ch$(W)=0$ do not.

As Fig.~\ref{WFphase1} reveals, the models $H_{\cal T}$ and $H_{\cal TV}$ have essentially
the same phase diagram with respect to the Chern number, although the topological state
is slightly larger in the $V$-shape potential, and an additional topological
state forms in the very small regime $F=0.38 \sim 0.4$ and $W<0.5$ in the system $H_{\cal TV}$.
[This small regime is system-size dependent as we show in appendix B.]
One may wonder if the shape of the potential may influence the value
of the Chern number through the boundary condition, but this is not the case.
We verified that all other physical quantities such as the level-spacing ratio
exhibit similar behavior in the both models.

\subsection{Edge modes and quasi-edge modes}\label{edgemode}

It is interesting to see spatial configurations of the spins 
(i.e., the local fermion densities).
This investigation is also useful to study the in-gap state, which is one of hallmarks
of topological phase.
To this end, we employ the open boundary condition (OBC) and
define $\Delta E_N\equiv E_{N+1}-E_N$, where $E_N$ is the 
ground state energy of the system with total $N$ 
up spins, [i.e., $(L-N)$ down spins, $m_z={N\over L}-{1 \over 2}$].
Hereafter, we use notation such that $N^\ast$ denotes $N$ for a topological state, i.e., a plateau
with $m_z={N^\ast\over L}-{1 \over 2}$. 
Similarly we introduce, $\Delta \rho_N(i)\equiv \rho_{N+1}(i)-\rho_N(i)$, where 
$\rho_N(i)=\langle\psi_N|S^z_i|\psi_N\rangle$ and $|\psi_N\rangle$ is a
many-body wave function of state with $N$ up spins. 
In the JW-fermion picture, $\Delta \rho_N(i)$ represents a density configuration of 
a single excitation particle on the topological plateau state with total $N$ particles.

\begin{figure}[t]
\centering
\begin{center}
\includegraphics[width=6cm]{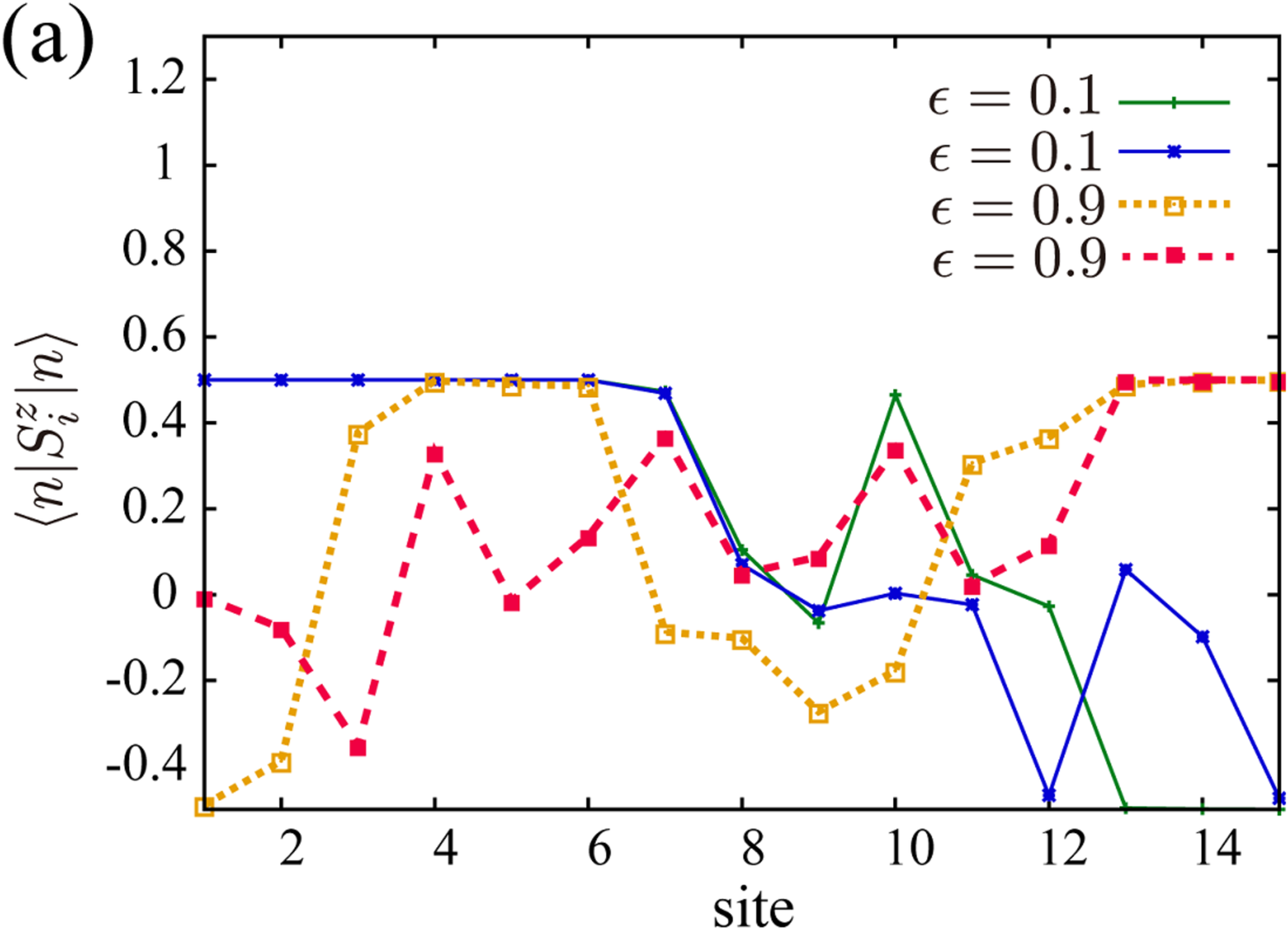}
\includegraphics[width=5.6cm]{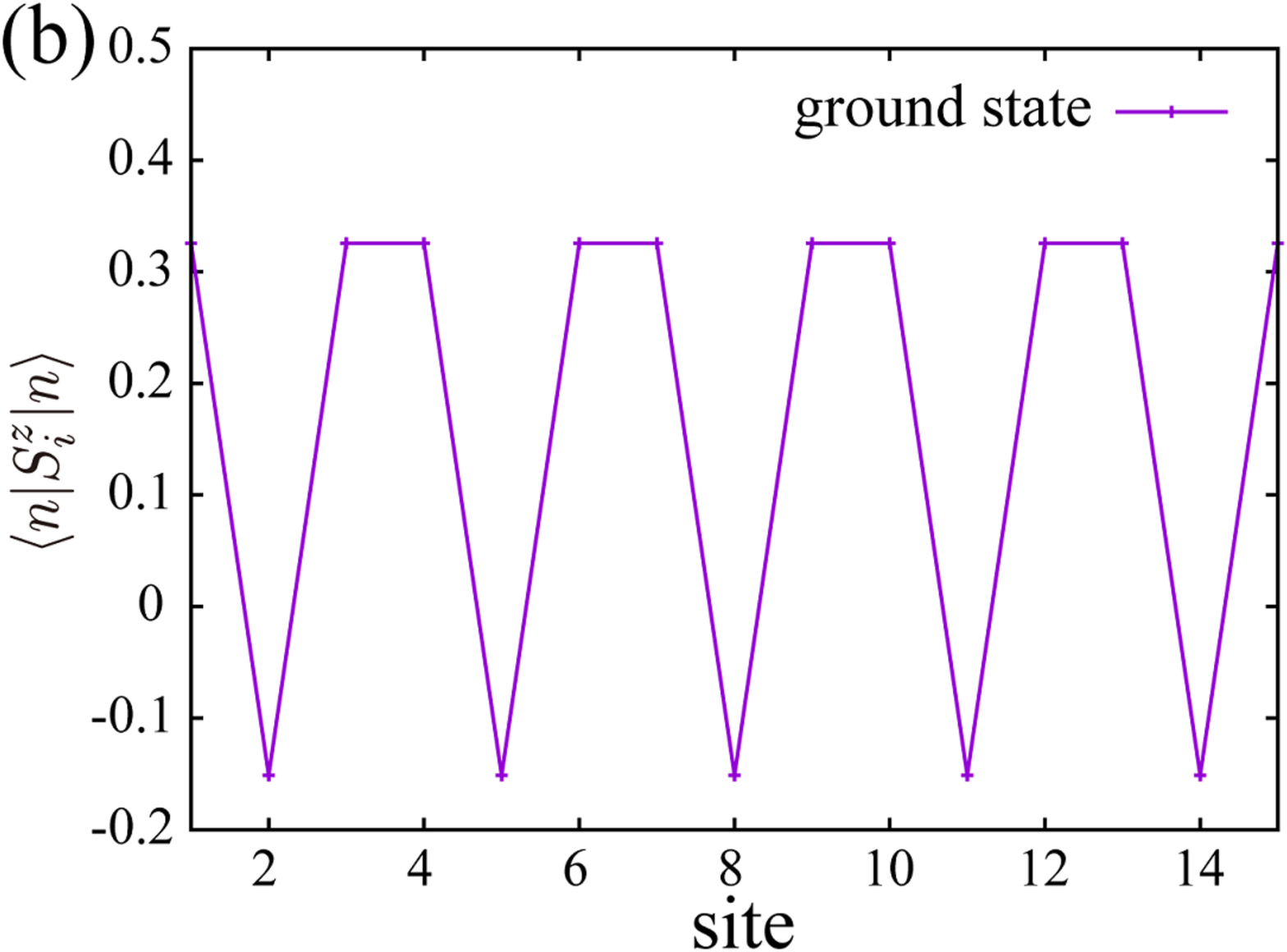}
\end{center}
\vspace{-0.5cm}
\caption{(a) Local-spin configurations $\rho_N(i)$ for $L=15, N=10$ and 
$F=1.0>F_{c1}, W=0.1$ with the open boundary condition. 
Two low and high-energy configurations are displayed that are located in the vicinity of
$\e=0.1$ and $\e=0.9$, respectively.
In the JW-fermion picture, 
the low-energy (high-energy) states have large amplitudes in left (right) side
of the system, in which the linear potential is low (high).
(b) The calculation of $\rho_N(i)$ for the ground state of the system with
$L=15, N=10$ and $F=W=0$.
}
\label{conf1}
\end{figure}

In Fig.~\ref{conf1}, we first show $\rho_N(i)$ of low and high-energy states
for $F=1.0$, $W=0.1$ as well as that of the ground state in the $F=0$ system.
The results for large $F$ show that the up spins in the low (high)-energy states reside
on the low (high)-potential side as 
the linear potential dominates the $(S^z-S^z)$ anti-ferromagnetic interaction in the system.
{\em This is an essential feature of the WS localization in many-body system 
in the presence of the `inter-particle repulsion'.}

\begin{figure}[t]
\centering
\begin{center}
\includegraphics[width=6cm]{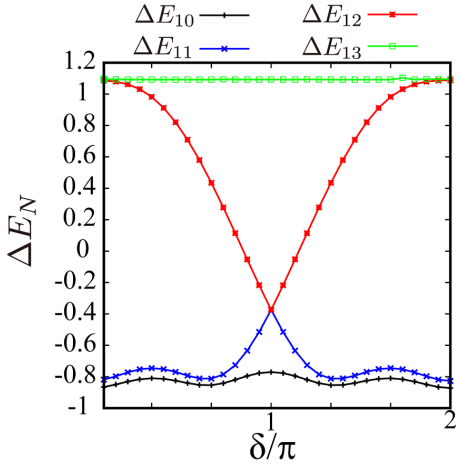}
\end{center}
\vspace{-0.5cm}
\caption{Energy differences $\Delta E_{10} \sim \Delta E_{13}$ for $F=W=0$.
$\D E_{12}$ and $\D E_{11}$ cross with each other at $\delta=\pi$.
This indicate that there exists a gapless edge mode at $\delta=\pi$.
$L=18$.
}
\label{edgemodeE1}
\end{figure}
\begin{figure}[h]
\centering
\begin{center}
\includegraphics[width=5.5cm]{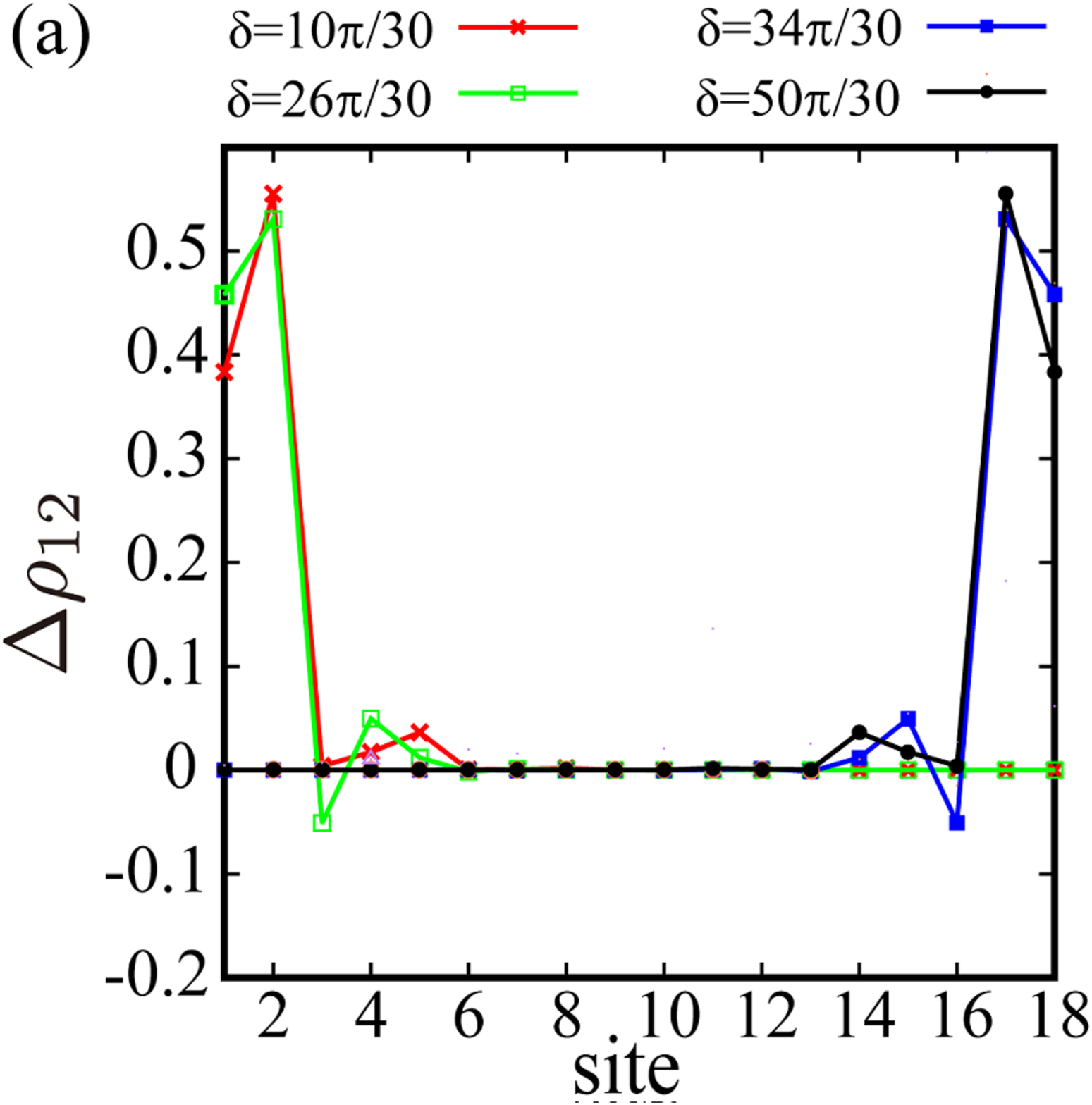}
\includegraphics[width=5.5cm]{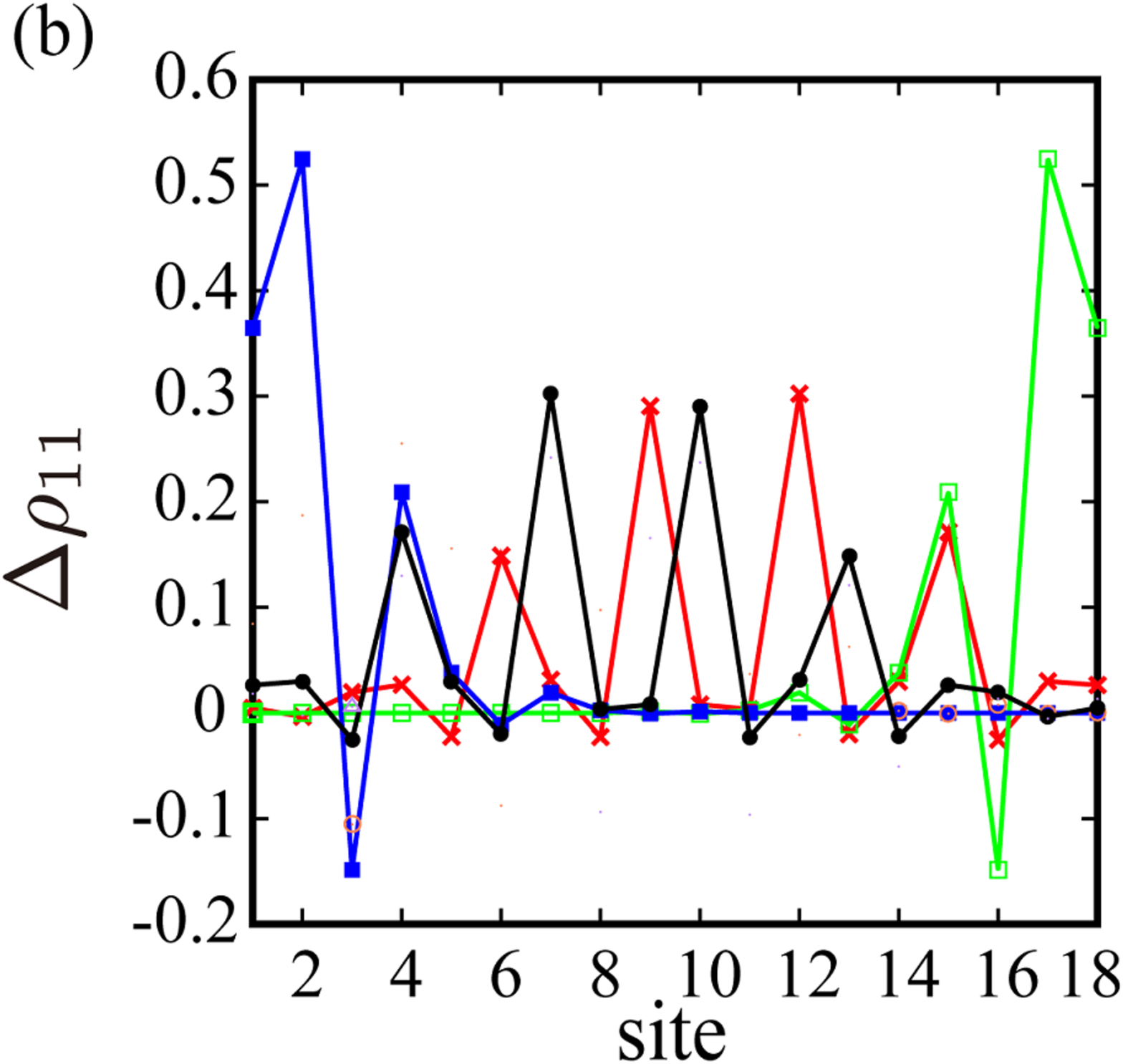}
\includegraphics[width=6.5cm]{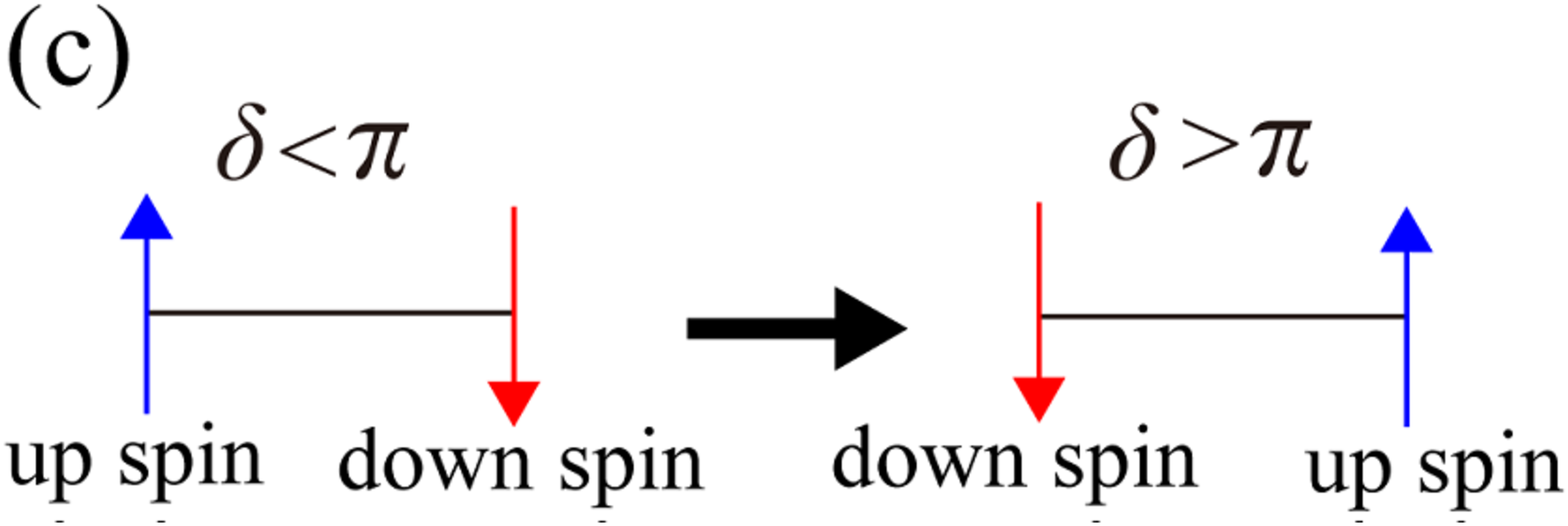}
\end{center}
\vspace{-0.5cm}
\caption{
(a) Edge modes $\Delta\rho_{12}$.
The edge mode in the upper band moves from left to right as $\delta$ increases.
(b) Edge modes $\Delta\rho_{11}$.
The edge mode in the lower band moves from right to left as $\delta$ increases.
With the results presented in Fig.~\ref{edgemodeE1}, these behaviors of $\D \rho_N$
indicate that a gapless edge mode exists at $\delta=\pi$ in the topological ground state.
$F=W=0$ and $L=18$.
(c) Schematic picture of edge excitations as $\delta$ varies.
Color of arrows referees to excitations in the JW-fermion picture shown in Fig.~\ref{band}.
}
\label{edgemode2}
\end{figure}

Let us turn to the edge modes existing in the topological ground state with the OBC.
We first focus on the case of $F=W=0$~\cite{Hu,Chen}.
By the ED, we obtain $\Delta E_N$ for $N=N^\ast$, $N=N^\ast\pm 1$ and $N=N^\ast -2$, and 
$\Delta\rho_N(i)$ for $N=N^\ast$ and $N=N^\ast -1$. 
The results are displayed in Figs.~\ref{edgemodeE1} and ~\ref{edgemode2} for $N^\ast=12$ 
with $L=18$.
As a function of $\delta$, $\Delta E_{N^\ast}$ has a minimum at $\delta=\pi$,
whereas $\Delta E_{N^\ast-1}$ has a maximum at $\delta=\pi$, and they cross 
with each other there.
On the other hand as seen in Fig.~\ref{edgemode2}, 
$\Delta\rho_N(i)$ for $N=N^\ast=12$ and $N=N^\ast-1=11$ exhibit a sharp peak at
left (right) and right (left) edges for $\delta \lesssim\pi \ (\delta\gtrsim\pi)$, respectively.  
These results indicate that the edge mode exists above the ground state of $N=12$, and
it is almost gapless in the vicinity of $\delta=\pi$
and exhibits sharp modulation of the spin amplitude at both edges.
In fact, the excess energy of the edge mode, $E_{\rm edge}$, is estimated by~\cite{Chen}
\be
E_{\rm edge}&=&\Delta E_{N^\ast}-\Delta E_{N^\ast-1}  \nonumber \\
&=&E_{N^\ast+1}+E_{N^\ast -1}-2E_{N^\ast}.
\label{EN}
\ee
Similarly, the local excess of up and down spins (nothing but the JW fermion density modulation), 
$\rho_{\rm edge}$, is given by
\be
\rho_{\rm edge}&=&\Delta \rho_{N^\ast}-\Delta \rho_{N^\ast-1}  \nonumber \\
&=&\rho_{N^\ast+1}+\rho_{N^\ast -1}-2\rho_{N^\ast}.
\label{rhoN}
\ee
Therefore, $E_{\rm edge}= 0$ at $\delta=\pi$ in the present case, and $\rho_{\rm edge}$
exhibits sharp plus and minus peaks at edges.

In order to verify the above conclusion and characterize the edge mode in the system, 
by cutting the system into two halves, we calculated an entanglement spectrum~\cite{Sirker} 
as a function of $\delta$. 
The topological phase with edge mode relates to the degeneracy of the lowest entanglement
spectrum of the ground state wave function.
We found that the ground state is four-fold degenerate at $\delta=\pi$ and 
this indicates the existence of a free spin-1/2 edge mode, i.e., the signal of the topological phase.  
On the other hand, the other states with higher energies do not.
We investigated other cases with various $\alpha$ and $N^\ast$, and found similar results.
The left-right edge modes interchange at the crossing point $\delta=\pi$ 
in the present case by reflecting ch$(W=0)=-1$.
On the other hand, there are multiple crossing points for higher ch$(W=0)$ cases.

\begin{figure}[t]
\centering
\begin{center}  
\includegraphics[width=5cm]{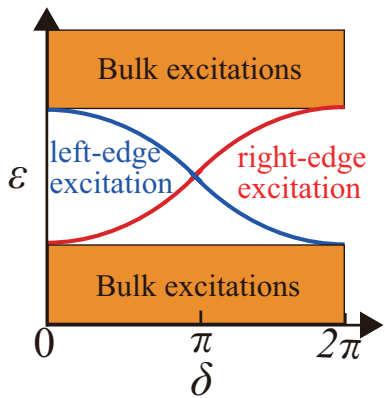}
\end{center}
\caption{Schematic picture of in-gap levels in {\em the single-body picture of the fermion}
in topological state with the open boundary condition.
Left-edge mode in the upper band and right-edge mode in the lower band cross with each other
at $\delta=\pi$ as $\delta$ increases.
Filled (empty) state of fermion corresponds to up-spin (down-spin) state.
}
\label{band}
\end{figure}

The above result can be understood by considering a non-interacting counterpart
of the modulated Heisenberg model in Eq.~(\ref{HPM}), i.e., the modulated XY model,
\be
H_{\rm XY}=\sum_i J_i (S^x_i S^x_{i+1}+S^y_i S^y_{i+1}).
\label{HXY}
\ee
This model is equivalent to the following non-interacting spinless fermion 
model connected by the JW transformation,
\be
H_{\rm fermion}={1 \over 2}\sum_i J_i (c^\dagger_ic_{i+1}+\mbox{h.c.}),
\label{Hfermion}
\ee 
where $c_i \ (c^\dagger_i)$ is fermion annihilation (creation) operator.
Schematic behavior of the single-particle spectrum in $H_{\rm fermion}$ [Eq.~(\ref{Hfermion})]
is shown in Fig.~\ref{band}.
Fermion topological state corresponds to the state in which $N=N^\ast$ fermions just fill
up the lower bulk bands.
In the single-particle picture,
the lowest-energy state in the upper band decreases its energy as $\delta$
increases until $\delta=\pi$.
On the other hand, the highest-energy state in the lower bands
increases its energy as $\delta$ increases to $\pi$.
Then, these two modes interchange at $\delta=\pi$.
The edge mode above the topological ground state is composed of the above left-right
single-particle edge modes in the XY-spin system, i.e.,
$c^\dagger_{\rm left}c_{\rm right}|\psi_G\rangle \ (c^\dagger_{\rm right}c_{\rm left}|\psi_G\rangle)$
for $\delta<\pi \ (\delta>\pi)$, 
where $|\psi_G\rangle$ is the topological ground state for $N=N^\ast$, and 
$c^\dagger_{\rm left} \ (c^\dagger_{\rm right})$ is the creation operator of the left (right) edge mode.
Our numerical calculations obviously indicate that the above single-particle
picture survives in the present interacting case.
We should also remark here that the system in Eq.~(\ref{Hfermion}) is directly related to
quantum Hall state (QHS) in a two-dimensional lattice.
As explicitly shown in Ref.~\cite{QHS},
the parameter $\delta$ in the modulated exchange $J_i$ in the system
Eq.~(\ref{Hfermion}) corresponds to the wave number $k_y$ and $\alpha i$ to
the vector potential in the $x$-direction. 
Then, the one-dimensional system Eq.~(\ref{Hfermion}) is extended to a two dimensional system. 
The band spanned by the momentum space $(k_x, k_y)$ is topological, where each band is characterized by own Chern number.
Therefore, the behavior of the edge mode in the spin system, which has a vanishing energy gap
at finite $\delta$, can be understood from the QHS point of view in which gapless edge modes
carry a transverse electric current.

\begin{figure}[t]
\centering
\begin{center}
\includegraphics[width=5.5cm]{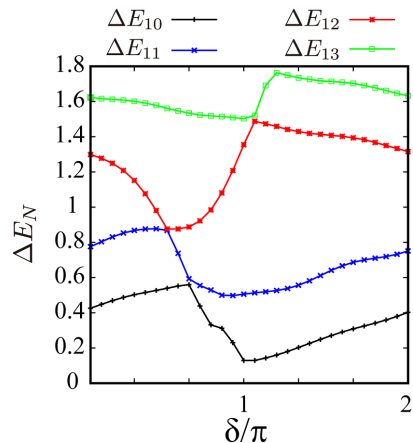}
\end{center}
\vspace{-0.5cm}
\caption{Energy differences $\Delta E_{10} \sim \Delta E_{13}$ for $F=0.1$ and $W=0$.
The calculations show that $\D E_N$'s cross with each other at three values of
$\delta=\delta_i \ (i=1,2,3)$.
In the single-particle picture, this indicates that energy crossing takes place there with a
vanishing energy gap.
$L=18$.
}
\label{edgemodeE2}
\end{figure}
\begin{figure}[t]
\centering
\begin{center}
\includegraphics[width=5.5cm]{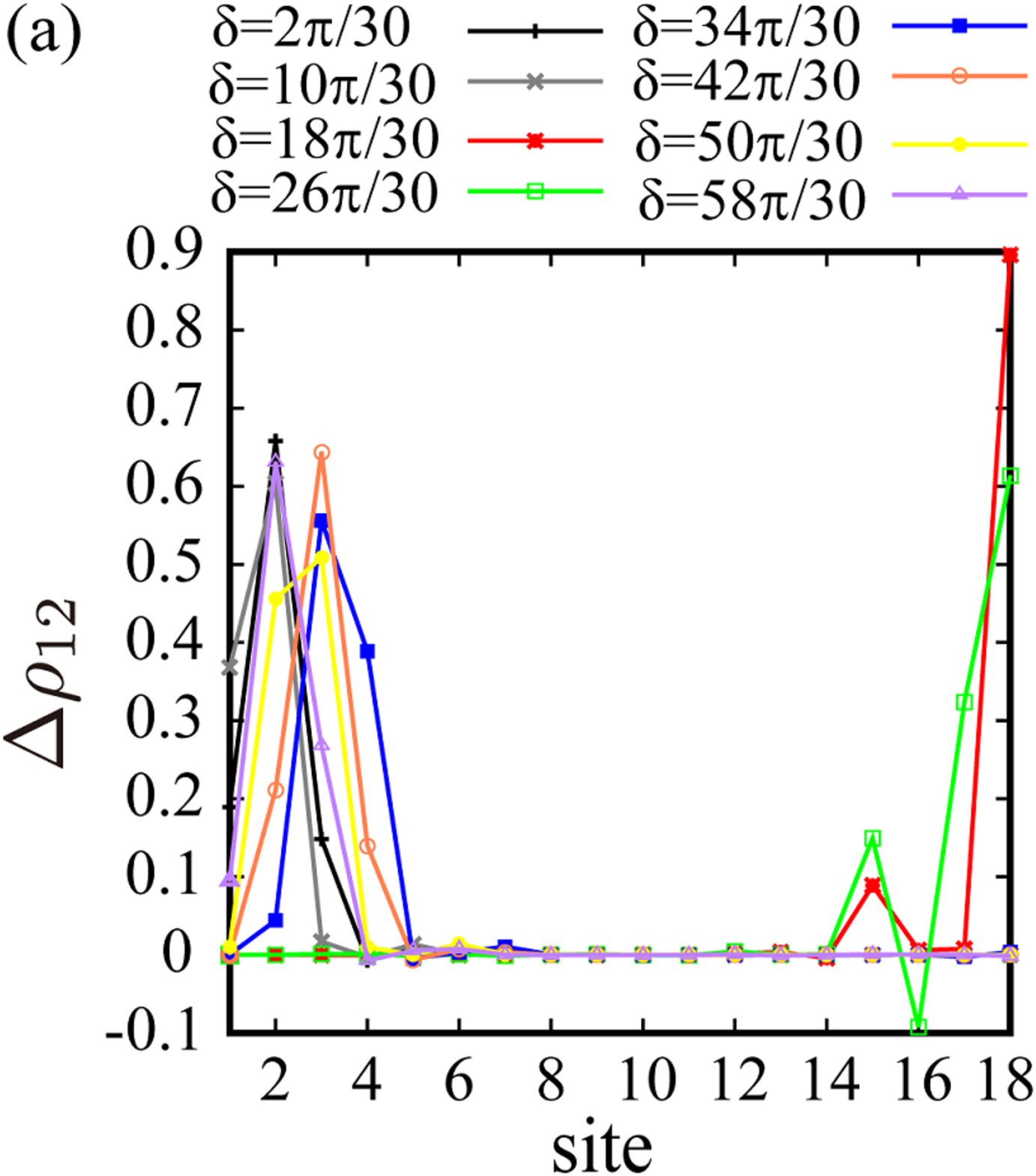}
\includegraphics[width=5.5cm]{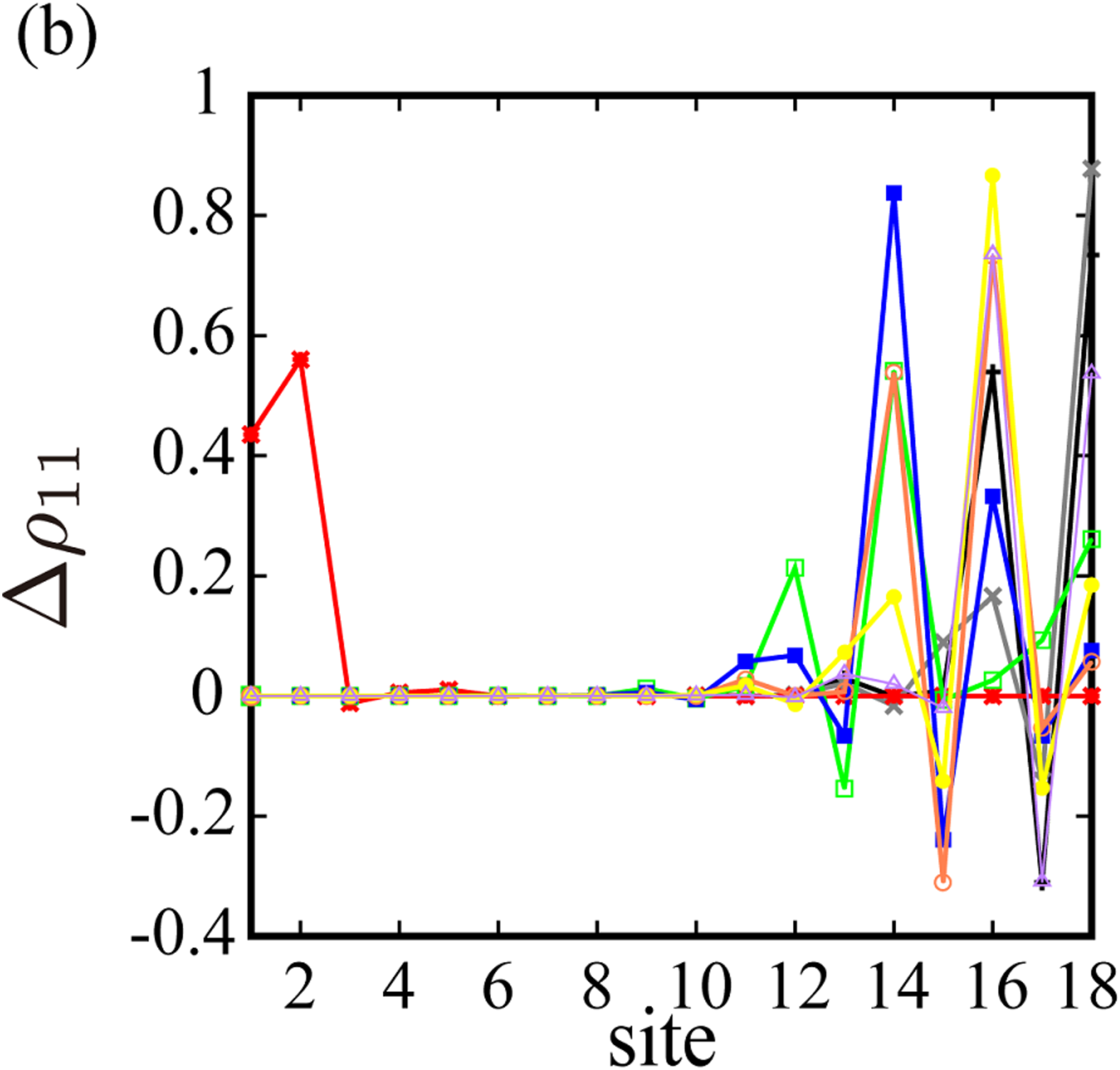}
\includegraphics[width=6.5cm]{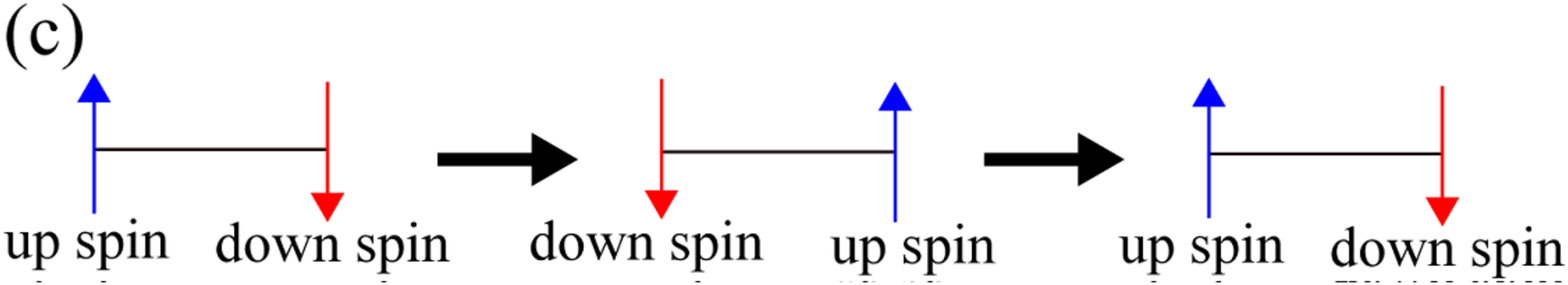}
\end{center}
\caption{(a) Edge modes $\Delta\rho_{12}$,
(b) Edge mdes $\Delta\rho_{11}$ for $F=0.1$ and $W=0$.
The peaks move from left to right and from right to left as $\delta$ varies.
Some of peaks represent the secondary modes that take the place of the genuine
edged modes of the $F=0$ case shown in Fig.~\ref{edgemode2}.
The state exists in topological phase, and we call the above modes quasi-edge modes.
(c) Schematic picture of edge excitations as $\delta$ increases.
$L=18$.
}
\label{edgemode3}
\end{figure}

Let us turn to the target system in the linear gradient potential.
In Figs.~\ref{edgemodeE2} and \ref{edgemode3}, 
we show the calculations of $\Delta E_N$ and $\Delta\rho_N(i)$ for various
values $\delta$ for $F=0.1$.
The energy differences $\Delta E_N$ exhibit different behavior from that of the $F=0$ case. 
$\Delta E_{11}$ and $\Delta E_{12}$ touch with each other at $\delta \simeq {8 \over 15}\pi$.
Besides that, $\Delta E_{12}$ touches with $\Delta E_{13}$ at $\delta \sim  {16 \over 15}\pi$,
and $\Delta E_{11}$ with $\Delta E_{10}$ at $\delta  \sim  {11 \over 15}\pi$.
In addition to this complicated behavior  of  $\Delta E_N$'s,
we observe very interesting behavior of $\Delta\rho_N(i)$ as the value of $\delta$ varies.
For example in $\D \rho_{12}$, the peak representing the edge mode first moves from left to right 
and stays there between $\delta={18 \over 30}\pi$ and $\delta={26 \over 30}\pi$.
After this stay, the peak returns to left for $\delta \geq {34 \over 30}\pi$.
This behavior is obviously directly related to the energy crossing of $\D E_N$'s observed in 
Fig.~\ref{edgemodeE2}.
That is, as $\delta$ increases, 
$\D E_{12}$ starts with $\D E_{12}(\delta=0)\simeq 1.3$ and crosses with $\D E_{11}$ at 
$\delta\simeq {16 \over 30}\pi$.
Then, $\D E_{11}$ crosses with $\D E_{10}$ at $\delta \simeq {32 \over 30}\pi$.
Similar behavior is seen for  $\D \rho_{11}$, and that is understood by  $\D E_N$'s.
The genuine profile of the local magnetization (i.e., fermion-density excess and deficiency) 
is given by Eq.~(\ref{rhoN}).
Careful look at the $\D E_N$'s in Fig.~\ref{edgemodeE2} reveals the fact that 
$\D E_{12}(\delta=0)\simeq \D E_{12}(\delta =2\pi)$ and 
$\D E_{11}(\delta=0)\simeq \D E_{11}(\delta =2\pi)$. 
This implies that the states with the second mode for $\delta < 2\pi$ 
smoothly transfer to those with the genuine edge mode for $\delta >0$.
[See the discussion below.]

The above phenomenon results from the gradient potential, i.e., in addition to the energy ramp 
in the classical picture, the localization tendency of quantum state plays an essential role.
Second highest-energy state
of the lower band also exhibits localized nature by the WS localization.
Second lowest-energy state in the higher band is similarly localized.
They have almost the same features in the real space with the edge modes. 
{\em Above calculations of $\Delta E_N$ and $\D \rho_N$ indicate that these secondary modes
take the place of the genuine edge modes for certain parameter regimes of $\delta$. 
Then, we call them quasi-edge modes.}
The same result to the above can be obtained by directly calculating the energy
of the first excited state above the topological ground state~\cite{Chen}.
However, the above analysis helps us to understand what actually happens.

As we mentioned in the above, the present model is related to the QHS, and $\delta$ 
corresponds to a component of wave vector, e.g., $\delta \sim k_y$.
The gapless edge modes contribute to the transverse conductivity under a magnetic field
that is determined by $\alpha$.
The emergence of quasi-edge modes in the present model implies that similar gapless excitations
exist in the QHS in a strong electric field and contribute to the transverse conductivity.
In other words, localized states via the WS mechanism might be observed in the QHS.
Experiments on QHS forming in a two-dimensional lattice can examine this prediction, 
and they are certainly welcome.

\begin{figure}[t]
\centering
\begin{center}
\includegraphics[width=5.5cm]{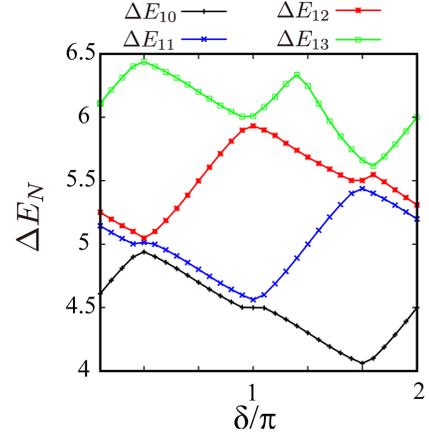}
\end{center}
\vspace{-0.5cm}
\caption{Energy differences $\Delta E_{10} \sim \Delta E_{13}$ for $F=0.5$ and $W=0$.
They exhibit complicated behaviors.
$L=18$.
}
\label{edgemodeE3}
\end{figure}
\begin{figure}[t]
\centering
\begin{center}
\includegraphics[width=6cm]{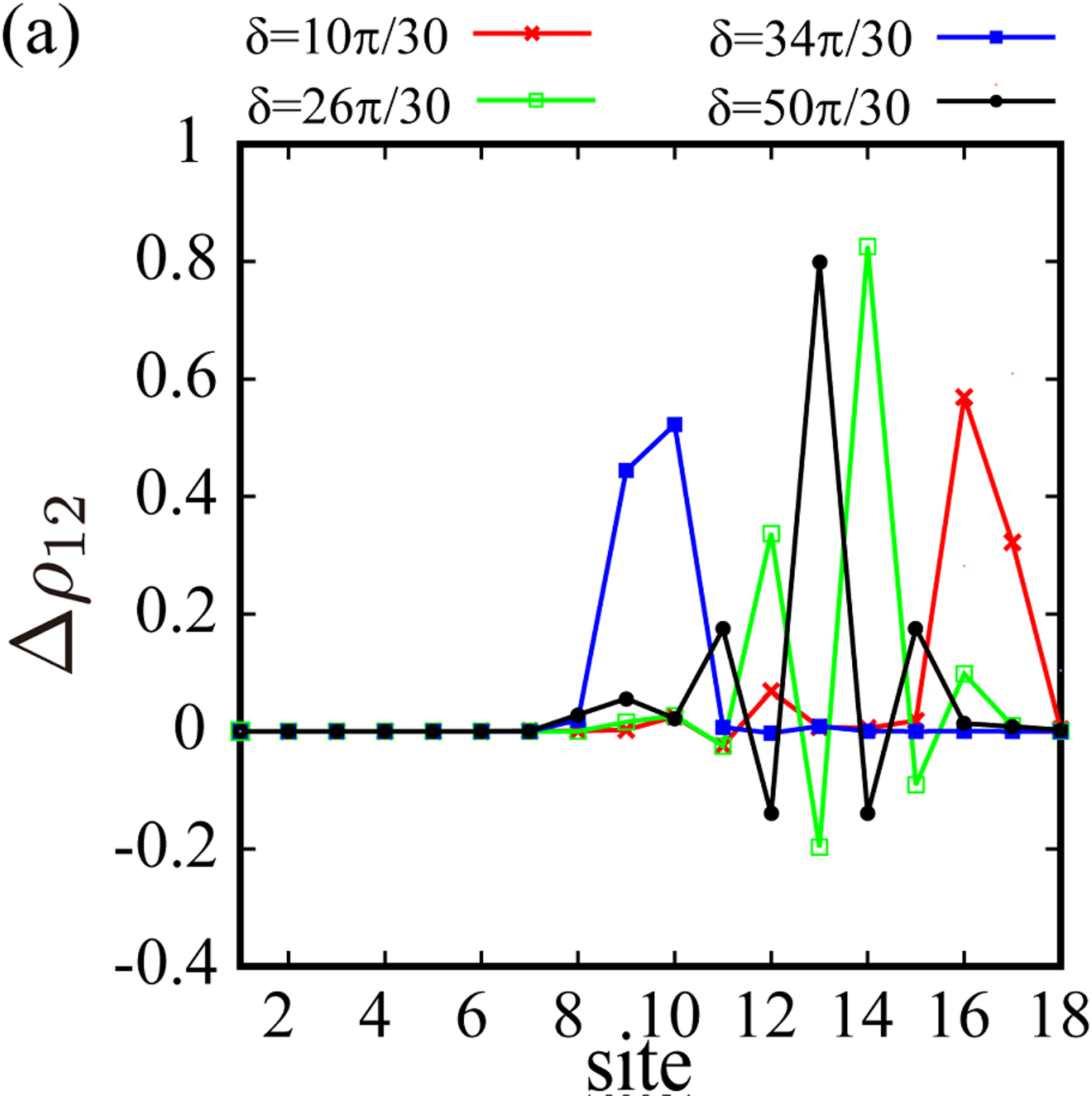}
\includegraphics[width=6cm]{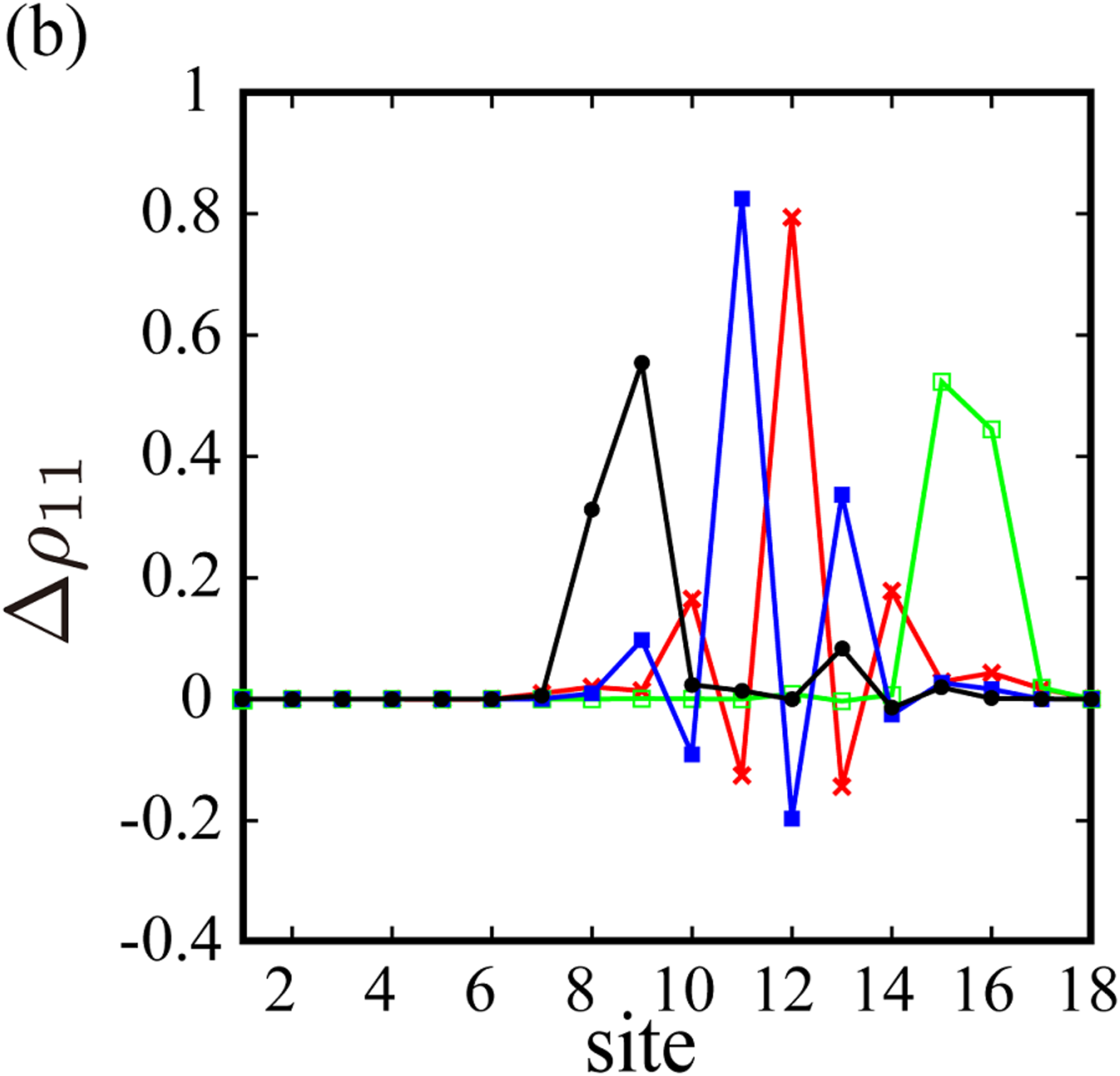}
\end{center}
\vspace{-0.5cm}
\caption{(a) Excitations shown by $\Delta\rho_{12}$,
(b) Excitations shown by $\Delta\rho_{11}$ for $F=0.5$ and $W=0$.
By the strong linear potential, topological state is destroyed and edge modes do not exist.
$L=18$.
}
\label{edgemode4}
\end{figure}

The calculation of the Chern number in Fig.~\ref{CHN1} indicates that the topological state 
is destroyed by the gradient potential for $F\gtrsim 0.2$.
Then, we investigate the system for $F=0.5$. 
The energy differences and density profiles are displayed in Figs~\ref{edgemodeE3} and \ref{edgemode4}.
$\D E_N$'s exhibit very complicated behavior as $\delta$ varies and some of them have almost 
the same value for finite regimes of $\delta$.
This means that there are no energy gaps in these regimes.
Also, the behavior of the energy differences does not exhibit energy crossings but avoided 
crossings instead. 
This is contrast to the result in Fig.~\ref{edgemodeE2}. 
This indicates no moving edge mode as $\delta$ varies.
On the hand, Fig.~\ref{edgemode4} shows that $\D \rho_N$ for $N= 11$ ans $12$ are located 
in the right part of the system and are obviously different from those in Fig.~\ref{edgemode3}.
This result can be understood intuitively, i.e., the linear gradient potential with $F=0.5$ is strong enough, 
and the left part of the system is filled with `particles' (up spins) first, and only right part
is empty for $N\simeq 10$ as the system size $L=18$.
This is the physical picture of the topological-state destruction by the gradient potential
in the present system.

\begin{figure}[t]
\centering
\begin{center}
\includegraphics[width=5cm]{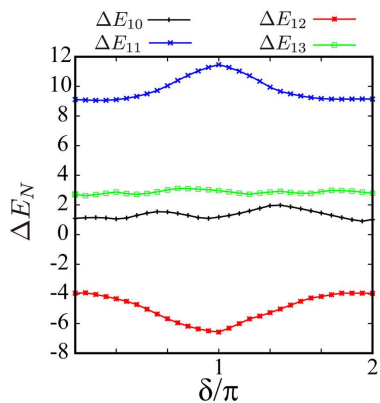}
\end{center}
\vspace{-0.5cm}
\caption{Energy differences $\Delta E_{10} \sim \Delta E_{13}$ for $F=0$ and $W=6$.
They have no level crossings.
$L=18$.
}
\label{edgemodeE4}
\end{figure}
\begin{figure}[h]
\centering
\begin{center}
\includegraphics[width=6cm]{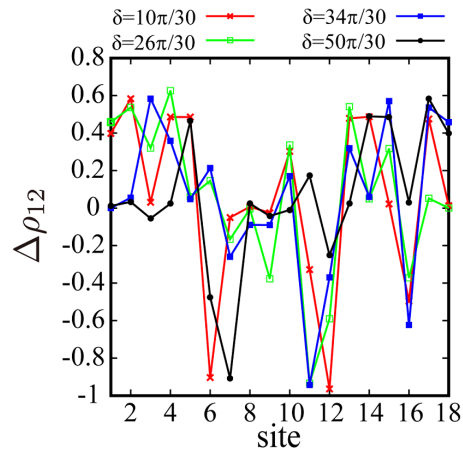}
\end{center}
\vspace{-0.5cm}
\caption{Density difference $\Delta\rho_{12}$, for $F=0$, $W=6$ and a specific disorder
realization.
They are dominated by the disorder pattern.
Similar profile is obtained for $\Delta\rho_{11}$.
$L=18$.
}
\label{edgemode5}
\end{figure}

Finally, we study the effect of the random potential $W$.
As Fig.~\ref{WFphase1} shows, the topological state is destroyed by the random potential
for $W\gtrsim 3$.
We investigate the system for $F=0$ and $W=6$, and the obtained results are displayed in 
Figs~\ref{edgemodeE4} and \ref{edgemode5} for certain specific $\{ h_i\}$.
No crossing takes place in $\D E_N$'s as $\delta$ varies. 
$\D \rho_N$'s fluctuate rather randomly due to the random potential 
and edge modes cannot be recognized.
This feature of the breaking of the topological state is quite different from that by the gradient potential.
Therefore, we expect that there is a crossover regime separating between the linear potential-breaking
and on-site disorder-breaking topological phase.
This point will be mentioned in Sec.~\ref{MBL}, in which MBL is studied.


\section{Transition to MBL in gradient potential}\label{MBL}

\subsection{Level-spacing statistics}\label{LSS}

In this section, we shall study a transition from an ergodic phase to
MBL phase by increasing $F$ and/or $W$.
The previous work~\cite{Refael} investigated the uniform coupling case of $\lambda=0$, and 
obtained the phase boundary such as $F_{c2}(W\to 0)\simeq 1.8$ and 
$W_{c2}(F=0)\simeq 4.0$. 
To obtain the MBL phase boundary of the present model, we first employ 
the level-statistics analysis as in Ref.~\cite{Alet,Refael}.
To this end, we first calculate energy-level spacings $\delta_n=E_{n+1}-E_n$ for a
fixed realization of the disorder and then obtain level-spacing ratios 
$r_n=\langle\mbox{min} (\delta_n,\delta_{n+1})
/\mbox{max} (\delta_n,\delta_{n+1})\rangle$, where $\langle \cdots \rangle$
denotes the average over disorder realizations of $\{h_i \}$.
In the ergodic state, the level statistics reveals the Wigner-Dyson (WD) distribution
for which $r_n \simeq 0.53$, whereas in the localized state, Poisson distribution
with $r_n=\ln 4-1\simeq 0.386$.
In this work, we show $r_n$ as a function of the rescaled energy $\epsilon$,
$r(\epsilon)$, and obtain a phase boundary in the $(F-\e)$ plane.
In the practical calculations, we consider 1800 energy levels around $\e$
and also 30 realizations of disorder. 

\begin{figure}[h]
\centering
\begin{center}
\includegraphics[width=5.5cm]{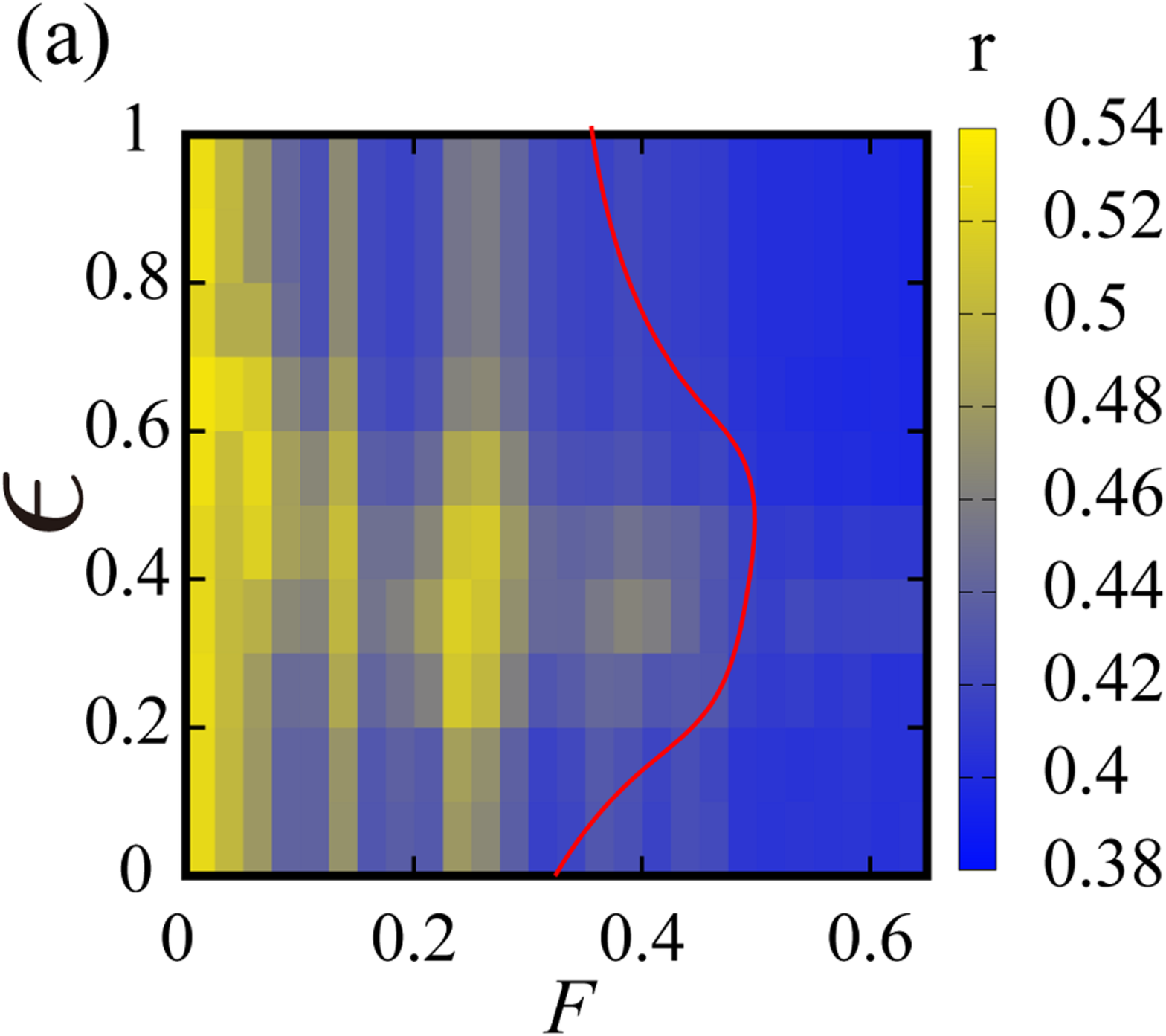} 
\includegraphics[width=5.5cm]{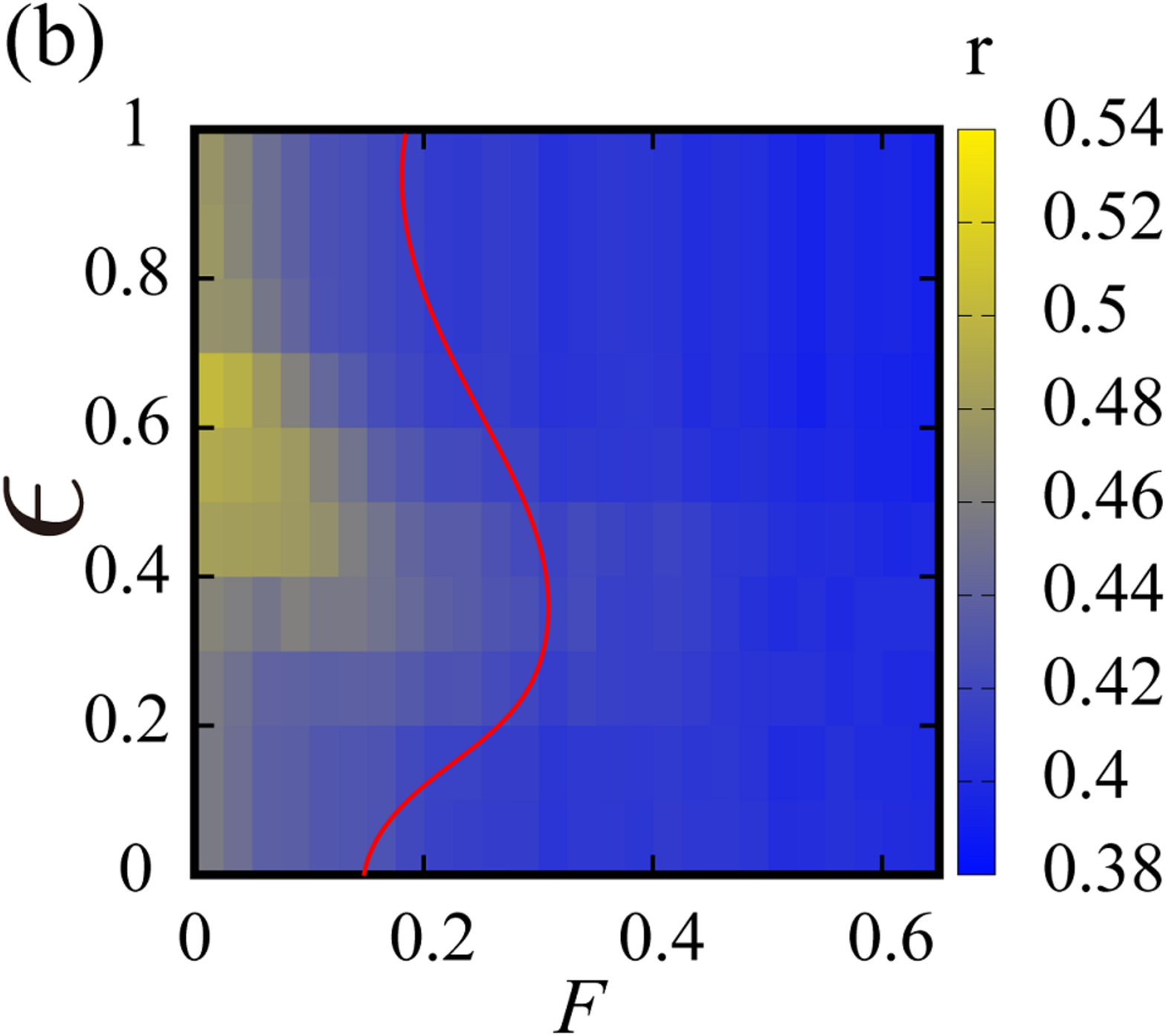}
\end{center}
\caption{(a) Level-spacing ratio as a function of the rescaled energy $\epsilon$
for very weak disorder $W=0.05$.
For an ergodic state $r\simeq 0.53$ and for a localized state $r\simeq 0.386$.
There are bands of extended states at $0.12\lesssim F\lesssim 0.15$ and $0.21\lesssim F\lesssim 0.28$.
This revival of the extended states is verified by calculating variance of the entangle entropy as shown
in Fig.~\ref{DStarget}.
30 realizations of the weak disorder were taken, and the data were obtained
by averaging across them.
(b) $r(\e)$ for $W=0.4$. Global structure of the phase diagram is the same
with that of (a), but bands of extended states do not exist.
}
\label{FWr}
\end{figure}

We consider the system with the parameters $N=12, \ L=18$ in which the topological 
ground state exists for $F=W=0$.
In Fig.~\ref{FWr} (a), we show the calculation of $r(\epsilon)$ as a function of $F$ with
very weak disorders $W \ll 1$, which was introduced to avoid the degeneracies
coming from the symmetries of the system without disorder, $W=0$.
It is obvious that $r(\e)$ decreases as $F$ is increased, and this behavior 
starts at smaller $F$'s in the low and high-edge regimes of the energy spectrum. 
However curiously enough, there exist band structure of extended states for 
$0.12\lesssim F \lesssim 0.15$ and $0.21\lesssim F\lesssim 0.28$.
Similar results are obtained for other system sizes such as $L=15 \ (12)$ and $N=10 \ (8)$.
Possible physical picture (origin) of these extended states is explained 
from the view of the Bloch oscillation and the Landau-Zener tunneling 
in Sec.~\ref{extendband}.
For moderate and strong disorder, bands of extended states do not exist,
as shown in Fig.~\ref{FWr} (b). 

The data in Fig.~\ref{FWr} (a) show that $r(\e)$ changes from the WD to Poisson
statistics at $F\sim 0.3$ in the low-energy states, besides the three regimes of extended state
mentioned above.
In Sec.~\ref{topology}, we observed that the system loses its topological nature 
at $F_{c1}=0.2$, and from $F=0.2$ to $0.4$, the critical regime exits.
The above result of $r(\e)$ seems to be correlated with the change of the ground state properties.
In other words, the low-energy states exhibit similar behavior with the ground state
with respect to the level statistics.
[In Sec.~\ref{topology}, we explained that the excited states do not have topological properties.]
In Fig.~\ref{FWr}, we draw the phase boundary $F_{c2}(\e)$ with the red curve by 
the condition such as $r(\e)=0.46$~\cite{Refael}.

\begin{figure}[t]
\centering
\begin{center}
\includegraphics[width=7cm]{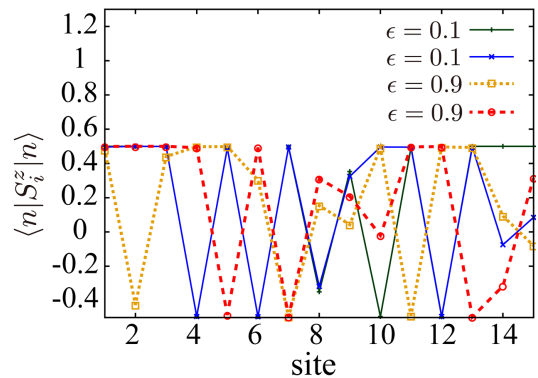}
\end{center}
\caption{Local spin configurations for $W=6$.
On-site random potential dominates the other terms in the Hamiltonian $H_{\cal T}$.
}
\label{spinconfW}
\end{figure}
\begin{figure}[t]
\centering
\begin{center}
\includegraphics[width=7cm]{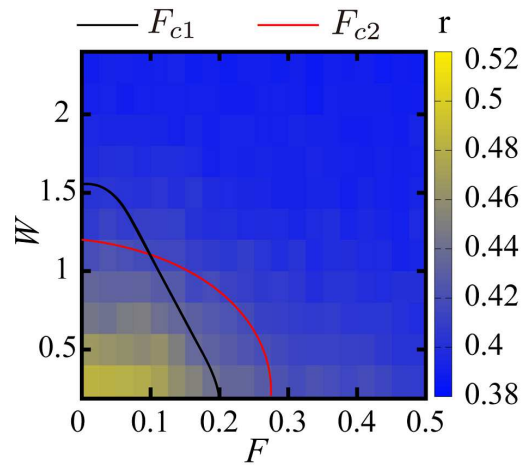}
\end{center}
\caption{Phase diagram of ergodic and MBL regimes for the Hamiltonian $H_{\cal T}$.
The phase boundary is determined by the value of $r(\e)=0.46$ in the central regime
of the energy spectrum.
The phase boundaries of the topological phase ($F_{c1}(W)$)
and the WS localization ($F_{c2}(W)$) are shown in the black and red lines, respectively.
$L=18$ and 30 realizations of the disorder.
}
\label{PDFW}
\end{figure}

In Fig.~\ref{conf1} in Sec.~\ref{topology}, we saw typical configurations of low 
and high-energy states in the regime $F=1.0>F_{c1}$.
By the above observation, the parameter in Fig.~\ref{conf1} is also located in the MBL regime.
Then, it is obvious that the low-energy states [$\e=0.1$] reside on the left side of 
the system, whereas the high-energy states [$\e=0.9$] the right side for 
$F=1.0$ and $W=0.1$.
This result indicates that the gradient potential dominates the other terms in the Hamiltonian
$H_{\cal T}$ [Eq.~(\ref{HT})] in the MBL state for $W \ll 1$.
On the other hand for the small gradient $F \ll 1$ and a strong disorder $W$,  
the spatial configurations $\{\rho(i)\}$ dominated by the disorder.
See Fig.~\ref{spinconfW}.

Phase diagram in the $(F-W)$ plane is displayed in Fig.~\ref{PDFW}.
The MBL phase boundary is determined by the value of $r(\e)$ in the central
regime of the energy spectrum.
In the regime $F<0.1$ and $1<W<2$, a coexistence phase of localization and 
topological order might form.
However, this is a consequence of variation of disorder realization $\{h_i \}$ as we explained 
in Sec.~\ref{edgemode}. 

\begin{figure}[t]
\centering
\begin{center}
\includegraphics[width=8cm]{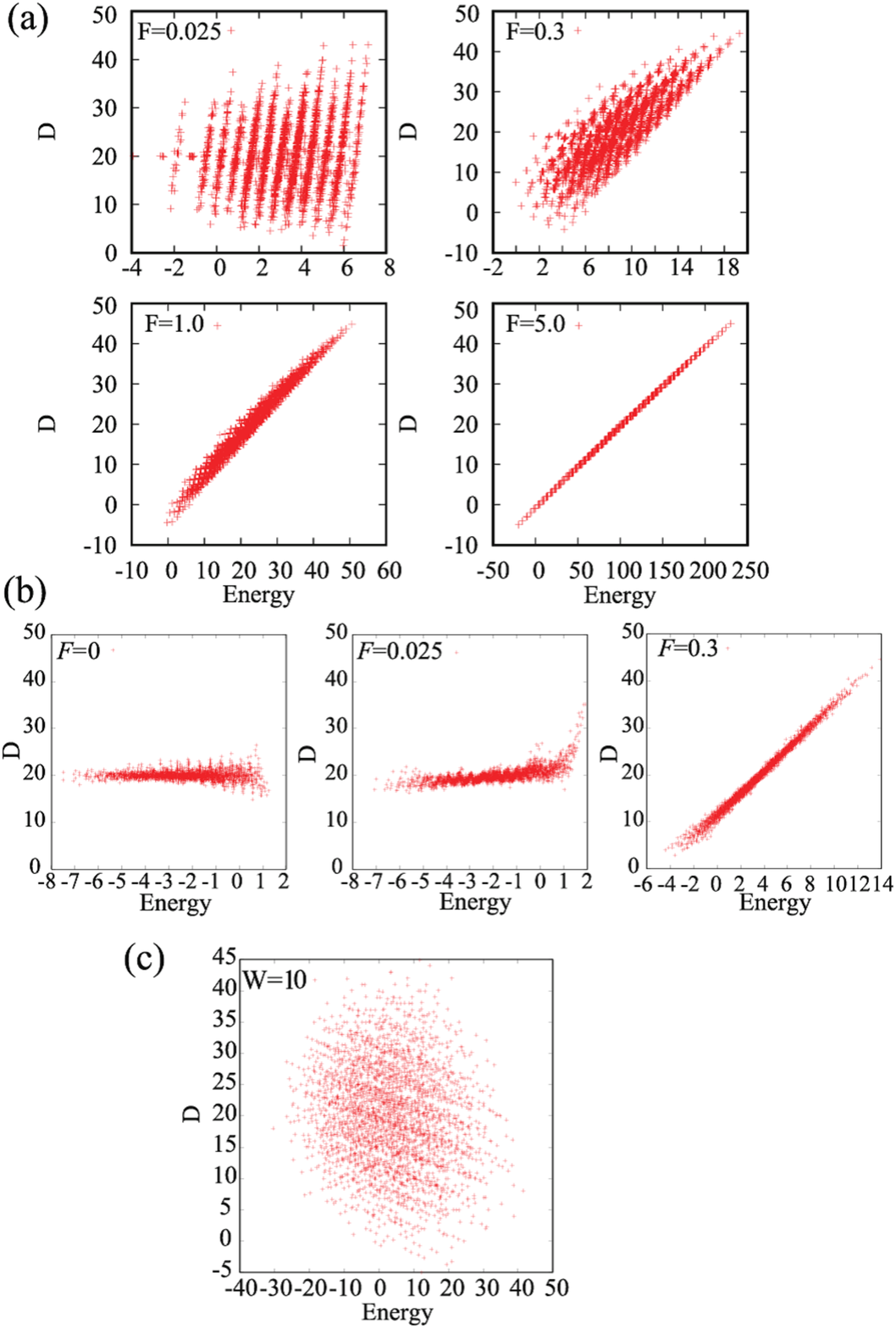}
\end{center}
\caption{(a)
Dipole moment $D$ as a function of energy for various values of $F$ with weak
disorder.
In the ergodic phase with $F=0.025$, the eigenstates in a given energy have a finite
spread in the dipole moment.
In the critical regime $F=0.3$, $D$ is getting large and the band structure becomes
obscure.
In the large $F$ case with $F=5.0$, the result indicates that 
the many-body wave functions have well-defined
dipole moment, which is large and linear to the energy. 
(b) $D$ for the uniform Heisenberg model ($\lambda=0$).
(c) Dipole moment for the case of strong disorder with $\lambda=0.8$.
Behavior is quite different from that of the weak-disorder case.
}
\label{dipole}
\end{figure}

\subsection{Dipole moment}

It is interesting to see how dipole moment defined by $D=\sum_iiS^z_i$ behaves
as a function of energy.
It gives us the global structure of the spatial configurations of the many-body states.
In Fig.~\ref{dipole} (a), we exhibit $D$ for various values of $F$ 
for the weak on-site disorder.
In the ergodic phase with $F=0.025$, the eigenstates in a given energy have a finite
spread in the dipole moment.
This behavior is similar to the band structure with a small band width,
and it is strong contrast to the case of $\lambda=0$~\cite{Refael}, 
where the average of $D$ is linear to the energy but there is no band structure. 
See Fig.~\ref{dipole} (b). 
In the critical regime $F=0.3$, the energy eigenvalue is getting large and the band structure becomes
obscure.
In the case of $F=1.0$, $D$ exhibits no band structure and is almost linear to the energy.
In the large $F$ case with $F=5.0$, the result indicates that 
the many-body wave functions have definite
dipole moment, which is large and linear to the energy reflecting the fact that
dynamics is severely restricted by the strong gradient potential.
On the other hand, in Fig.~\ref{dipole} (c), we show $D$ for $F=0$ and $W=10$.
The result shows that the dipole moment is randomly distributed as spin configurations are determined
by the strong disorder.
{\em Above result implies that there exists a crossover in the MBL phase, i.e.,
a crossover between the WS-localized regime with large $F$ and the on-site disorder 
dominant regime with large $W$}. 
In other words, there is a crossover between a translational symmetric MBL and the disorder induced conventional MBL.
In fact, this kind of crossover was recently discovered for a generalized Creutz ladder model~\cite{Creutz} and quantum simulation model of lattice gauge theory~\cite{Park}.

\begin{figure}[t]
\centering
\begin{center}
\includegraphics[width=8cm]{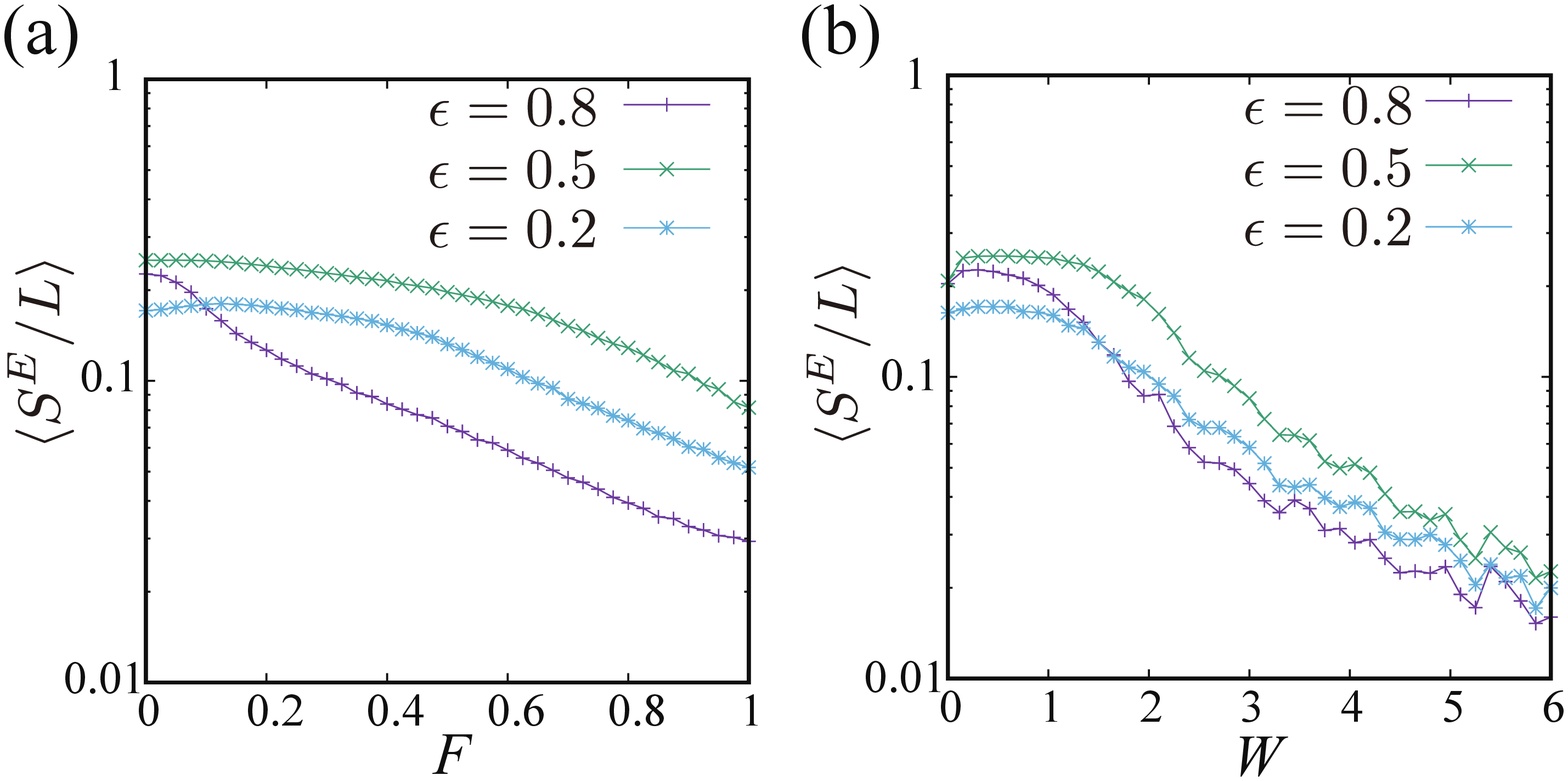}
\end{center}
\caption{(a) Entanglement entropy in the Heisenberg model as a function of $F$ for $W=0.3$.
(b) Entanglement entropy in the Heisenberg model as a function of $W$ for $F=0$.
}
\label{EE1}
\end{figure}

\subsection{Entanglement entropy}\label{EE}

Finally, we show calculations of the entanglement entropy, which is another hallmark of MBL.
We employ the von Neumann entanglement entropy, which is calculated
by dividing the system into subregion $A$ of size $L_A$ and $B$ of size
$L_B=L-L_A$, and define 
\be
S=-\mbox{Tr}_A(\rho_A\log \rho_A); \;\;\; 
\rho_A=\mbox{Tr}_B|\Psi \rangle \langle \Psi|,
\label{EES}
\ee
where $|\Psi\rangle$ is an exact eigenstate of the whole system, and 
$\rho_A$ is the reduced density matrix for the subregion $A$ obtained by tracing out
all the degrees of freedom of the complement subregion $B$.
As in the calculation of the level-spacing ratio $r(\e)$, we calculate $S$ 
as a function of the system energy, $S=S^E(\e)$.
In particular, we focus on the three cases $\e=0.2, 0.5$ and $0.8$ as typical cases, 
and take an average of $S$ evaluated for 50 eigenstates with energy $\simeq \e$.
Also for the on-site disorder, we average $S^E(\e)$ for 50 realization of $\{h_i \}$ for each $W$.
The total system size $L=15$ and $L_A=7, \ L_B=8$.

We first consider the case of $\lambda=0$, i.e, Heisenberg spin chain in the linear
gradient magnetic field.
In Fig.~\ref{EE1} (a), we show the calculations of $S^E(\e)$ as a function of $F$ for
fixed $W=0.3$.
As $F$ increases, all $S^E(\e)$s decrease rather rapidly.
In particular, $S^E(\e=0.8)$ is smaller than the others indicating that 
states in the high-energy regime tend to localize strongly by the gradient potential.
On the other hand, the disorder strength dependence of $S^E(\e)$ is shown in 
Fig.~\ref{EE1} (b), and we find
that $S^E(\e)$ tends to increase in the small-$W$ regime.
In the whole $W$ region, the states in the middle of the energy spectrum tend to extend
more compared to the states in the low and high regions.
This observation is consistent with the calculations of the level-spacing ratio
obtained in the previous work~\cite{Refael}.

\begin{figure}[t]
\centering
\begin{center}
\includegraphics[width=4cm]{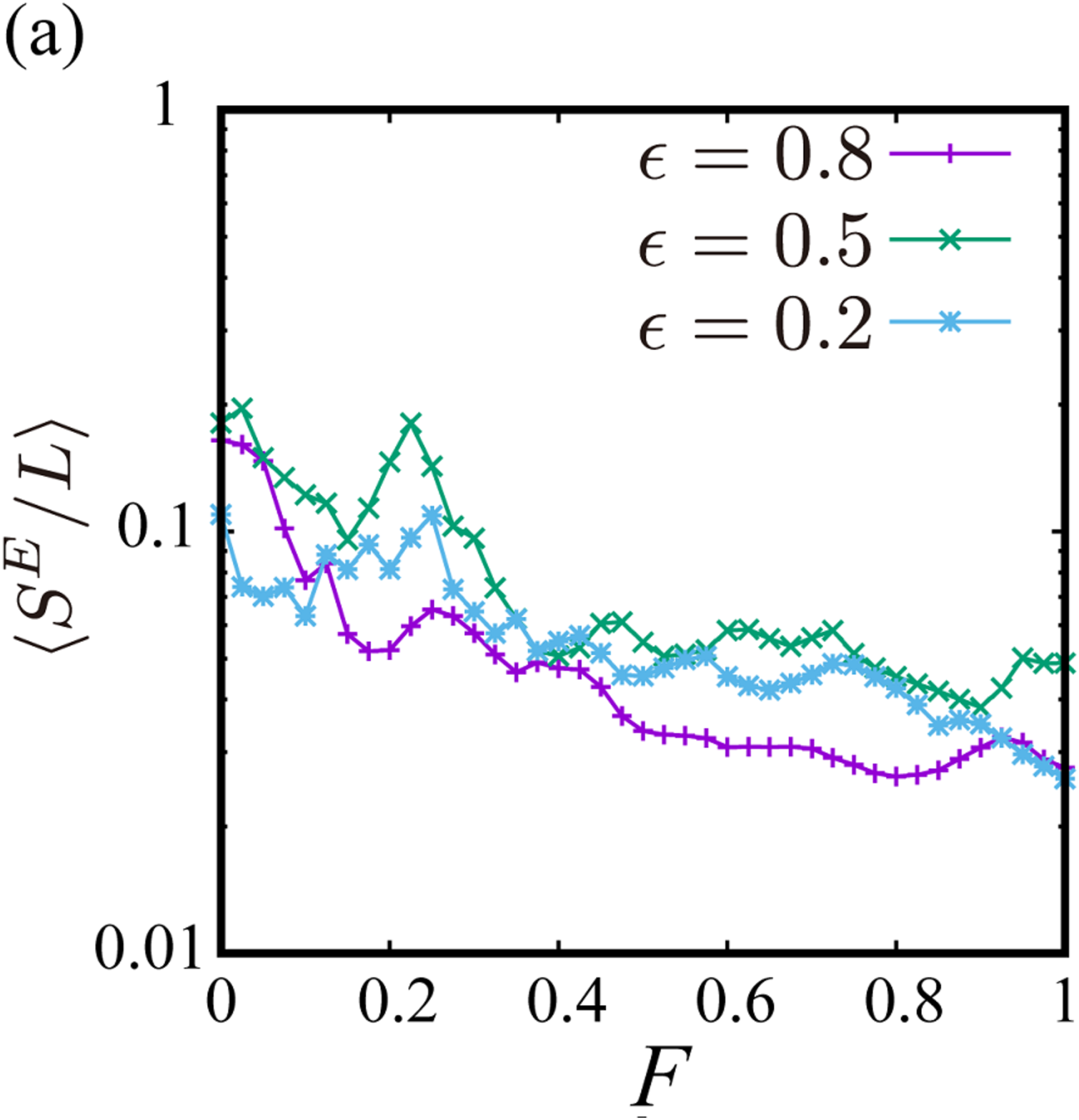}
\includegraphics[width=4cm]{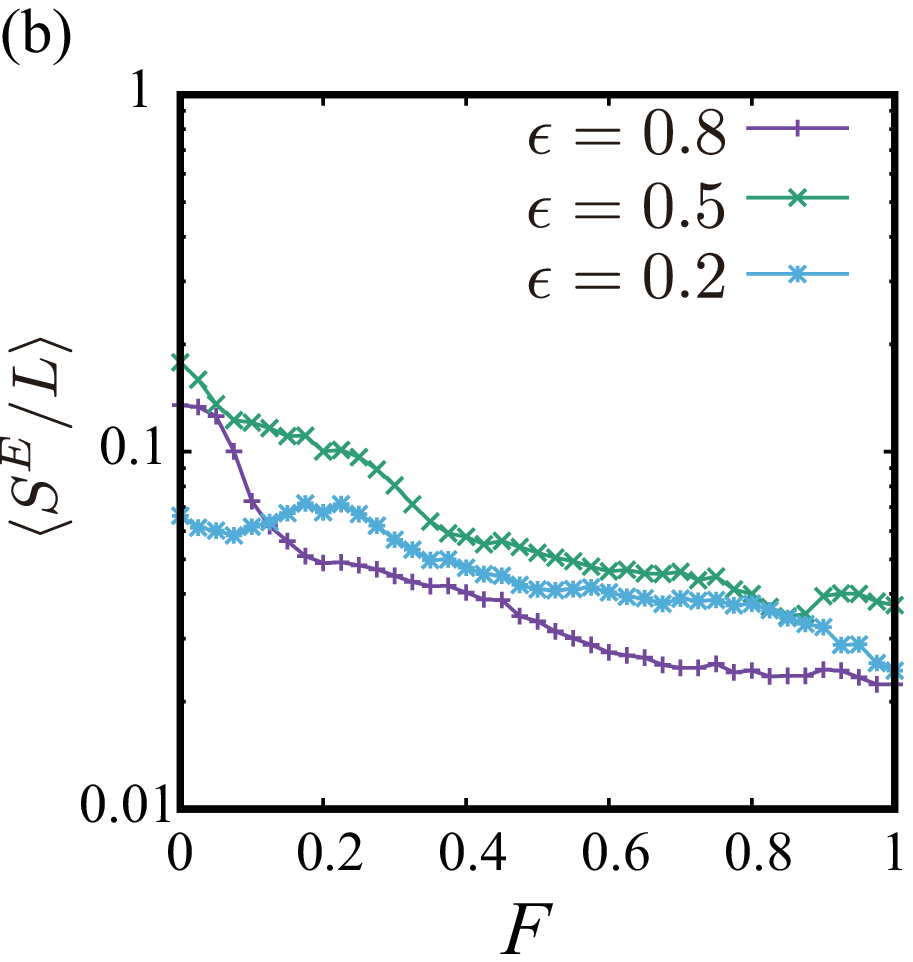}
\end{center}
\caption{Entanglement entropy in the present model as a function of $F$.
$W=0.05$ and $0.3$ in (a) and (b), respectively.
In case (a), the entanglement entropy exhibits sharp peaks for $0.2 \lesssim F\lesssim 0.28$.
In particular for the low and middle energies, $\e=0.2$ and $0.5$, this behavior is clear.
}
\label{EE2}
\end{figure}

Let us turn to the present model, $H_{\cal T}$ with $\lambda=0.8$. 
We first consider the case of a very weak disorder $W=0.05$ for which
the level-spacing analysis is shown in Fig.~\ref{FWr} (a).
Calculations of $S^E(\e)$ as a function of $F$ are presented in Fig.~\ref{EE2} (a).
Interestingly enough, Fig.~\ref{EE2} (a) shows that $S^E(\e)$ has a peak
in the regime $0.2 \lesssim F\lesssim 0.28$ indicating the existence of extended states there,
in particular for the low and middle energies, $\e=0.2$ and $0.5$, this behavior is clear.
This result is in good agreement with the calculation of $r(\e)$ in Fig.~\ref{FWr} (a),
which indicates the existence of extended states in that parameter regime as we explained
in Sec.~\ref{LSS}.
As $W$ is increased to $W=0.3$, this behavior disappears as seen in Fig.~\ref{EE2} (b).
We studied the case with $W=0.6$, and obtained the similar result. 
Then, it is obvious that the above interesting behavior stems from the interplay
of the modulated hopping $J_i$ and the gradient potential. 

Generally, $S^E(\e)$ is a decreasing function of $F$ for all $W$s.
This behaver is another evidence that dynamics of the system is severly restricted 
by the gradient potential and the WS localization takes place as $F$ is increased beyond $F_{c2}$.

In recent paper~\cite{Pollmann,Alet,Huse}, it was indicated that the standard deviation of 
the entanglement entropy can be used as a diagnostic for the ergodic to MBL transition or crossover.
We apply this diagnostic to the system studied in Fig.~\ref{FWr} (a) and Fig.~\ref{EE2} (a),
i.e., $W=0.05$.
To this end, we define two kinds of standard deviation of the half-chain entanglement entropy,
which we call sample-to-sample and eigenstate-to eigenstate deviations, respectively,
following Ref.~\cite{Huse}.
Definitions of them are as follows,
\be
&&\langle S\rangle \equiv {1 \over N_SN_E}\sum^{\rm sample}_i
\sum^{\rm state}_j S^j_i,  \nonumber  \\
&&\langle S\rangle_i \equiv {1 \over N_S}\sum^{\rm state}_j S^j_i,  \nonumber
\ee
\be
&&\Delta_{\rm SA}=\Big( {1\over N_SN_E}
\sum^{\rm sample}_i\sum^{\rm state}_j(S^j_i-\langle S\rangle)^2\Big)^{1/2}, \nonumber \\
&&\Delta_{\rm ET}={1\over \sqrt{N_S}N_E}\sum^{\rm sample}_i\Big(\sum^{\rm state}_j
(S^j_i-\langle S\rangle_i)^2\Big)^{1/2},
\ee
where $N_S \ (N_E)$ is the number of samples (eigenstates) used for evaluation, and
$S^j_i$ is the entanglement entropy of eigenstate $j$ in sample $i$.
Usually $\Delta_{\rm SA}>\D_{\rm ET}$, and for the case in which
fluctuations across samples are very small,  $\Delta_{\rm SA}\simeq\D_{\rm ET}$.
In the practical calculation, we consider 250 energy eigenstates
in the vicinity of $\e=0.5$~\cite{Huse} and 10 disorder samples for the $L=18$ target
system.

\begin{figure}[t]
\centering
\begin{center}
\includegraphics[width=4.2cm]{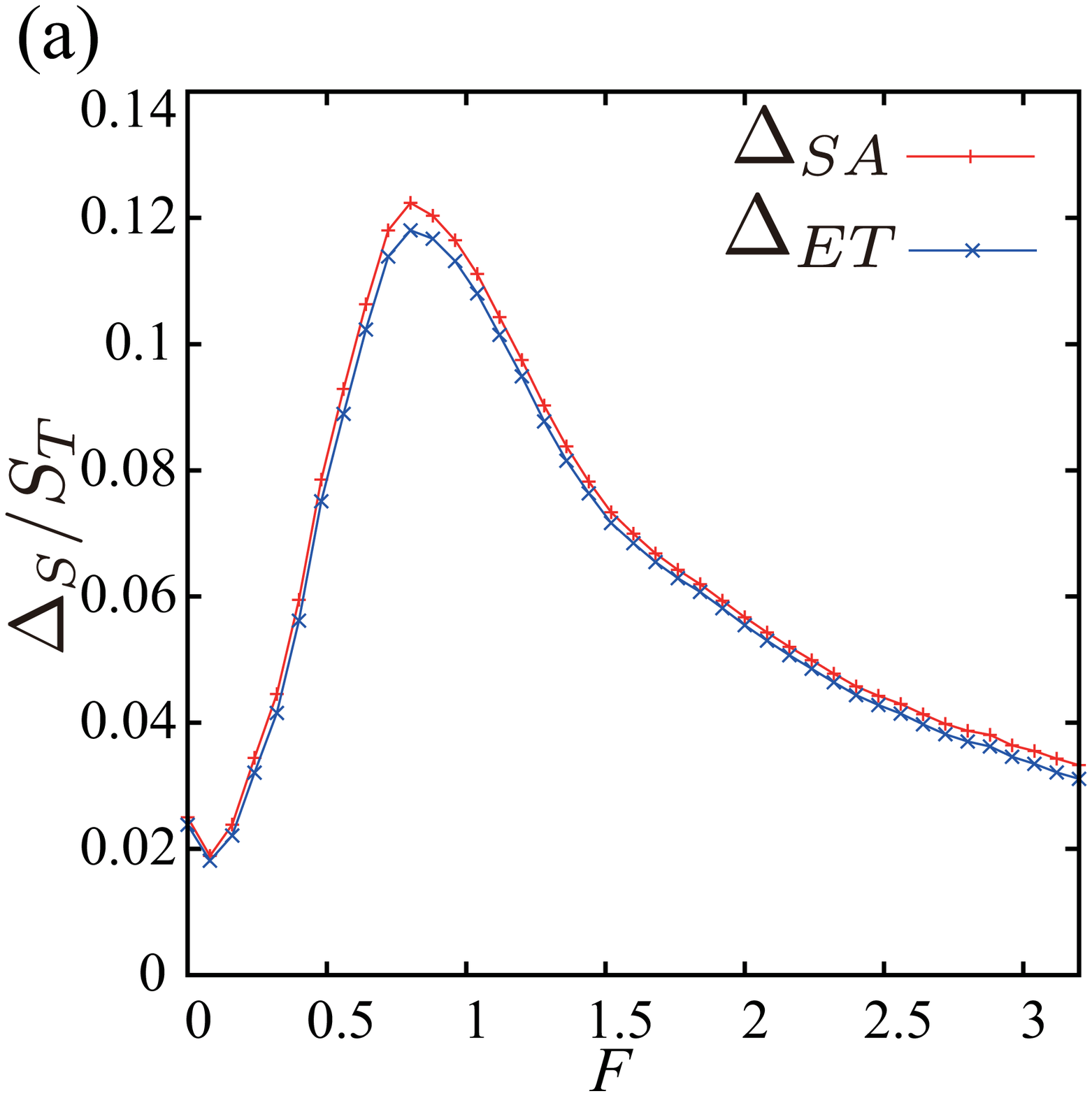}
\includegraphics[width=4.2cm]{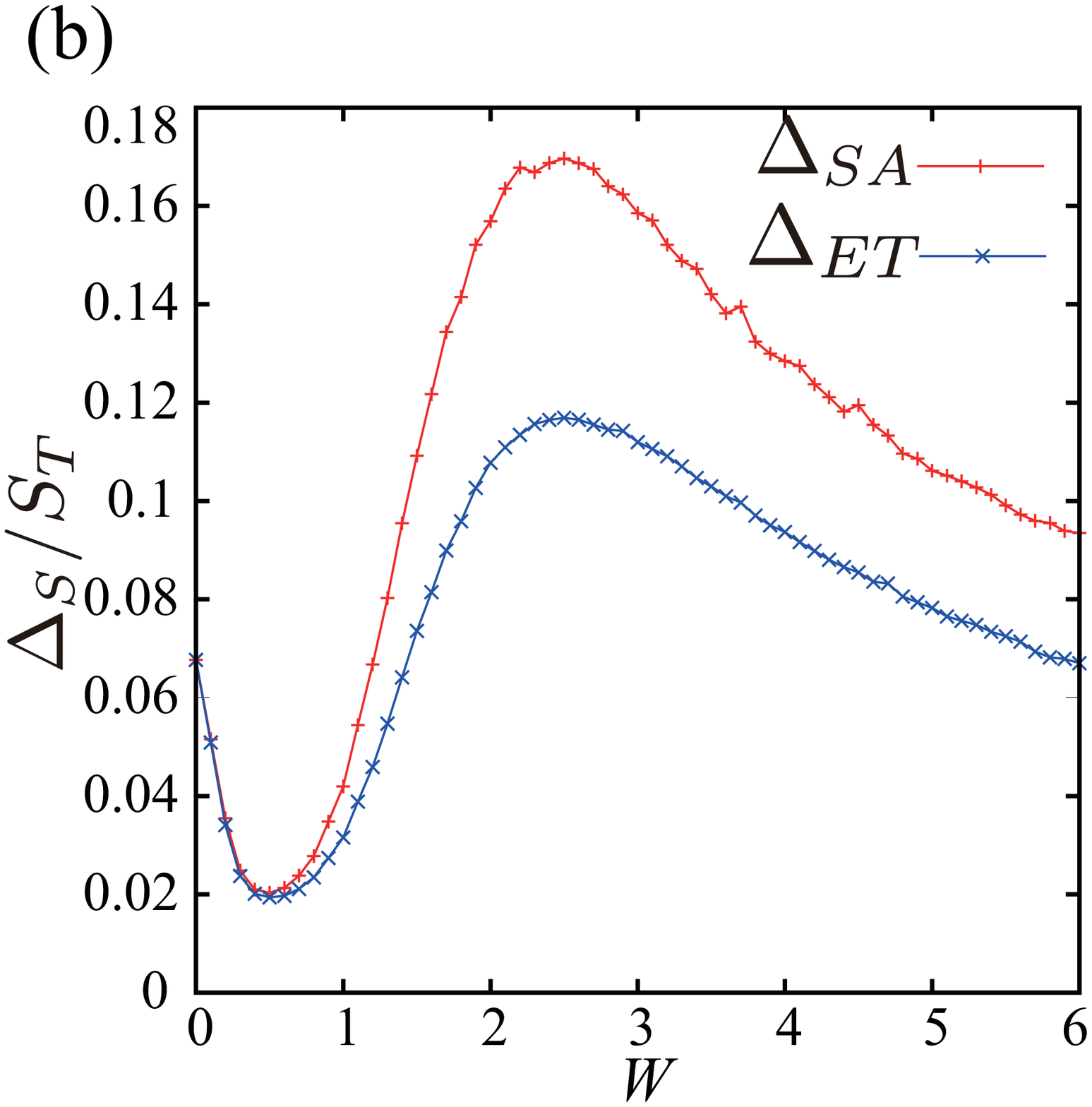}
\end{center}
\caption{Standard deviation of entanglement entropy in the Heisenberg model
corresponding to entanglement entropy in Fig.~\ref{EE1}.
(a) $\Delta_{\rm SA}$ and $\Delta_{\rm ET}$ as a function of $F$ for $W=0.3$.
(b) $\Delta_{\rm SA}$ and $\Delta_{\rm ET}$ as a function of $W$ for $F=0$.
$L=12$.
$S_T$ is the thermal entropy, $S_T=0.5(L \ln (2)-1)$.
}
\label{DSHeisenberg}
\end{figure}

We first show the results for the ordinary uniform Heisenberg model, i.e., $\lambda=0$ case.
In the previous paper~\cite{Refael}, the critical parameters were estimated as follows,
$F_{c2}(W=0)\simeq 0.9$ [from Fig.~1 of Ref.~\cite{Refael}] and 
$W_c(F=0)\simeq 3.75$~\cite{FN1}.
Figure~\ref{DSHeisenberg} displays $\Delta_{\rm SA}$ and $\Delta_{\rm ET}$ 
as a function of $F$ ($W$) for $W=0.3 \ (F=0)$.
For the case of $W=0.3$, $\Delta_{\rm SA}$ and $\Delta_{\rm ET}$ are very close with
each other.
This means that fluctuations caused by disorder samples is very small.
$\Delta_{\rm SA}$ and $\Delta_{\rm ET}$ exhibit a peak at $F\sim 0.8$, which obviously
corresponds to the ergodic to MBL crossover by the WS localization.
On the other hand for the case of $F=0$, we have $\Delta_{\rm SA}>\Delta_{\rm ET}$ as 
the ergodic to MBL crossover in this parameter regime stems from the disorder due to samples.
$\Delta_{\rm SA}$ and $\Delta_{\rm ET}$ exhibit a rather wide `peak' including the critical
value $W_c(F=0)\simeq 3.75$ obtained by the previous works~\cite{Alet,Wc1,Wc3,Wc4}.
However, $\Delta_{\rm SA}$ and $\Delta_{\rm ET}$ starts to increase at a rather smaller value 
of $W$ compared to the above value.
This means that the sample dependence of the critical value, $W_c$, is substantially large,
and calculations of large systems are needed to obtain an accurate critical value of $W$.

\begin{figure}[t]
\centering
\begin{center}
\includegraphics[width=8cm]{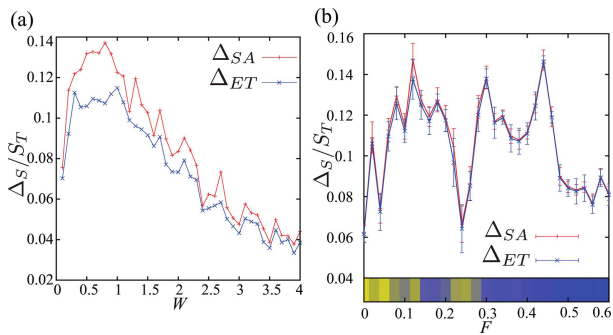}
\end{center}
\caption{Standard deviation of entanglement entropy in the target model
corresponding to Fig.~\ref{EE2}.
(a) $\Delta_{\rm SA}$ and $\Delta_{\rm ET}$ as a function of $W$ for $F=0$.
(b) $\Delta_{\rm SA}$ and $\Delta_{\rm ET}$ as a function of $F$ for $W=0.05$.
The color band in the bottom shows the level-spacing ratio in Fig.~\ref{FWr} (a) for $\e\sim 0.5$.
$L=18$.
Data points are average across 250 states and 10 disorder realizations.
$S_T$ is the thermal entropy, $S_T=0.5(L \ln (2)-1)$.
}
\label{DStarget}
\end{figure}

Let us turn to the target model.
We show the calculations in Fig.~\ref{DStarget}.
Fig.~\ref{DStarget} (a) displays $\Delta_{\rm SA}$ and $\Delta_{\rm ET}$ as a function of 
$W$ for $F=0$.
As in the uniform Heisenberg model in the above, both of them exhibit a rather wide `peak'
starting around $W\sim 0.4$ and ending around $W\sim 2$.
Obviously, this behavior of the variance of the entanglement entropy
corresponds to the ergodic and MBL crossover observed in the phase diagram in Fig.~\ref{PDFW}.
Fluctuations among disorder samples are fairly large in the above parameter regime.

The result in Fig.~\ref{DStarget} (b) gives detailed observation of the phase diagram displayed 
in Fig.~\ref{FWr} (a), and it shows very interesting results.
There are three large peaks in both $\Delta_{\rm SA}$ and $\Delta_{\rm ET}$, and
these peaks indicate that coexisting phase of extended and localized states forms there.
It is obvious that these peaks are closely related to the three bands of `extended states',
which we observed in Sec.~\ref{LSS}.
In fact, the dips of  $\Delta_{\rm SA}$ and $\Delta_{\rm ET}$ in Fig.~\ref{DStarget} (b)
indicate the locations in which delocalized state forms. 
This phenomenon is not observed in the uniform Heisenberg model, and also the strong
on-site disorder in the target system hinders the emergence of the bands of  the extended state.
Therefore, the interplay between the spatially-modulated exchange coupling 
and the gradient potential generates this unusual phenomenon.
In the following Sec.~\ref{extendband}, we shall briefly discuss possible origin of this phenomenon.

\subsection{Origin of the first breaking of the topological phase 
and the revival of the extended states for $W\ll 1$}\label{extendband}

In the previous subsection, the evaluation of $\Delta_{\rm SA}$ 
and $\Delta_{\rm ET}$ in Fig.~\ref{DStarget} certainly revealed their behaviors 
consistent with the phase diagram displayed in Fig.~\ref{FWr} (a).
We observed that the band structure with a revival of extended states appears on
the $(\epsilon$-$F$) plane, as verified by the peaks of $S^{E}$, $\Delta_{\rm SA}$ 
and $\Delta_{\rm ET}$ on increasing the value of $F$.

\begin{figure}[t]
\centering
\begin{center}
\includegraphics[width=7.5cm]{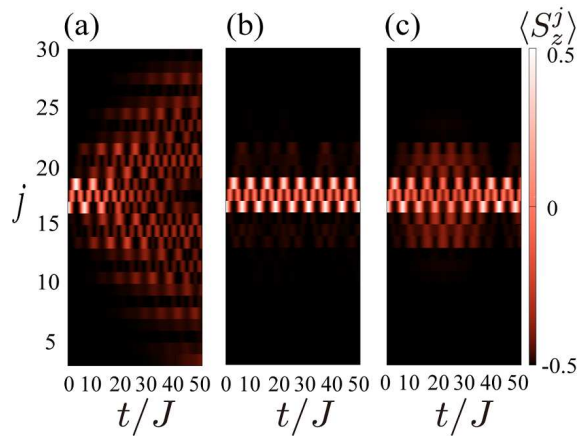}
\end{center}
\caption{Single particle dynamics: (a) $F=0$, (b) $F=0.1$, and (c) $F=0.3$. 
Single localized particle is put on a single site as an initial state. 
For case (a), after a short period, simple ballistic like expansion takes place.
For case (b), particle exhibits Bloch oscillation.
The amplitude of the oscillation is larger than the lattice spacing.
For case (c) with a stronger LP, particle tends to extend by a Landau-Zener mechanism.
}
\label{dynamics}
\end{figure}

We can explain the first destruction of the topological extended state at $F\sim 0.2$ from 
the single-particle picture of the model. 
To this end, let us consider noninteracting case of Eq.~(\ref{Hfermion}). 
Then the model has three bulk topological bands of the fermion corresponding to
the parameter $\alpha=1/3$~\cite{QHS}. 
As studied in the above, we consider the case in which the particle is filled nearly
up to the second band. 
There, applying the LP leads to a Bloch oscillation of a particle. 
For the Bloch oscillation, we can estimate the oscillation amplitude in the real space, 
which corresponds to the localization length. 
Actually, we numerically simulate a single particle dynamics by varying $F$ and obtain
the results shown in Fig.~\ref{dynamics} (a)$\sim$(c). 
The particle dynamics is certainly affected by the gradient of the LP potential, $F$.
Simple analytical calculation~\cite{Kolovsky}  shows that
the oscillation amplitude is given by $A=\frac{1}{F}\Delta E_{2}$, 
where $\Delta E_{2}(>0)$ is the band width of the second band in the present case.
If the half of the amplitude is less than the lattice spacing $d$, the particle is substantially
localized in single site and the topological phase is expected to be broken. 
From the actual value of $\Delta E_{2}$ and the condition $A/2 \lesssim d$, 
the single-site localization appears approximately for $F \gtrsim 0.15$, 
where the topological phase is destroyed.
As Fig.~\ref{dynamics} (b) shows, the single particle clearly exhibits localization dynamics. 
Certainly in Fig.~\ref{CHN1} for $F \gtrsim 0.15$, the topological phase vanishes 
and the WS localization tendency appears in Fig.~\ref{FWr} (a), although the calculations
in Sec.~\ref{MBL} include interactions.

In addition, we can argue the first revival of the extended state 
from the single-particle picture. 
In the Bloch oscillation, the states with a finite momentum in the second band also oscillate. 
For sufficiently large $F$, finite momentum states of the second band transit to 
the third band via avoided crossing between the second and third bands, i.e. 
the Landau-Zener transition~\cite{Landau,Zener}. 
We expect that since the third band can be regarded as a conduction band, the above
effect makes particle extended, i.e., the extended tendency is enhanced in the whole system. 
Numerical calculation can actually capture phenomena related to the effects mentioned above. 
As Fig.~\ref{dynamics} (c) for large $F$ indicates, the particle tends to extend more 
compared with the $F=0.1$ case in Fig.~\ref{dynamics} (b). 
This mechanism is reminiscent of a non-adiabatic breakdown of the Mott insulator in 
the Hubbard model under electric fields~\cite{Oka,Oka2}.
We also investigated whether a similar phenomenon takes place in the uniform 
Heisenberg model and found that the result is negative.

Finally, we point out the WS resonance as another possible explanation
of the observed revival of the extended states for increasing 
$F$~\cite{Sachdev,Buyskikh,Gluck,Para}, 
This phenomenon occurs only when some multi-band structure exists.
In our model, the model actually is in a multi-band situation due to the modulated coupling $J_{i}$
and the ($S^{z}-S^{z}$) interactions.  
Interplay between the interaction and the gradient potential may induce a resonance,
which hops a particle beyond unit cell and as a result, extended states appear.
Possibility of such a kind of resonance seems low, but it cannot be excluded.

\section{Conclusion}\label{conclusion}

In this paper, we systematically studied the interplay between the topological order and 
the WS localization in the modulated $s=1/2$ spin chain mainly by using the ED.
We first investigated the stability of the topological state against the LP, and 
estimated the critical gradient for destruction of the topological state.
Then, we numerically studied the in-gap excitations in the topological state 
to find that there appear the quasi-edge modes besides the genuine edge modes as a gapless excitation.
This is a precursor of the WS localization, which is to be observed in the topological state.

In the second half of the present work, we investigated localization by the LP as well as the diagonal
disorder, and obtained the phase diagram.
In this study, we found unexpected phenomenon, i.e., 
the revival of the extended states in the intermediate values of $F$.
Existence of this regime was verified by the calculation of the variance of the entanglement entropy.
The possible origin of this phenomenon was discussed, but the its complete understanding
is a future work.

The present model is feasible in experiments on cold atomic gases~\cite{Taylor} and we hope that 
our findings will be observed in experiments in the near future.
As we stressed in the text, the model is also closely related to the QHS on a two-dimensional lattice,
and the LP is nothing but a constant electric field applied to fermions.
Then, it is interesting to study two-dimensional electron systems in strong
 magnetic and electric fields.

Very recently, interesting idea of ``shattering'' and ``fragmentation'' of the Hilbert space 
by dipole-moment conservation was proposed~\cite{shattering1,shattering2,shattering3}.
In Sec.~IV.B, we observed that the dipole moment tends to a good quantum number for
large $F$s.
This indicates that MBL in the large-$F$ regime can be understood as a result of
the ``shattering''/``fragmentation'' phenomenon of the Hilbert space.
This is a future problem.

\bigskip
\section*{Acknowledgments}
Y. K. acknowledges the support of the Grant-in-Aid for JSPS Fellows (No.17J00486).


\appendix
\renewcommand{\thefigure}{\Alph{section}.\arabic{figure}}
\setcounter{figure}{0}
\renewcommand{\theequation}{A.\arabic{equation}}

\section{System-size dependence for Chern number calculation}\label{appendixA}
\begin{figure}[h]
\centering
\begin{center} 
\includegraphics[width=6cm]{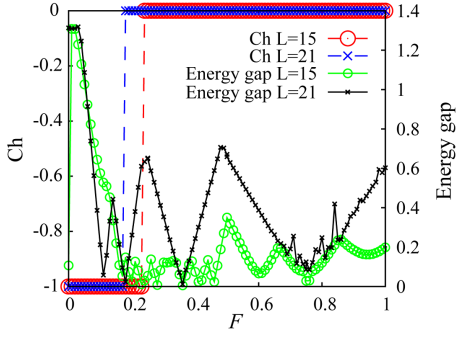} 
\end{center}
\vspace{-0.5cm}
\caption{Chern number and excitation gap as a function of $F$ calculated for
$L=15$ and $L=21$ systems.
Syetem-size dependence of the Chern number seems rather small.
On the other hand, the energy gap has a finite system-size dependence, in particular,
for $F>0.2$.
}
\label{systemsize1}
\end{figure}
\begin{figure}[h]
\centering
\begin{center}
\includegraphics[width=4.2cm]{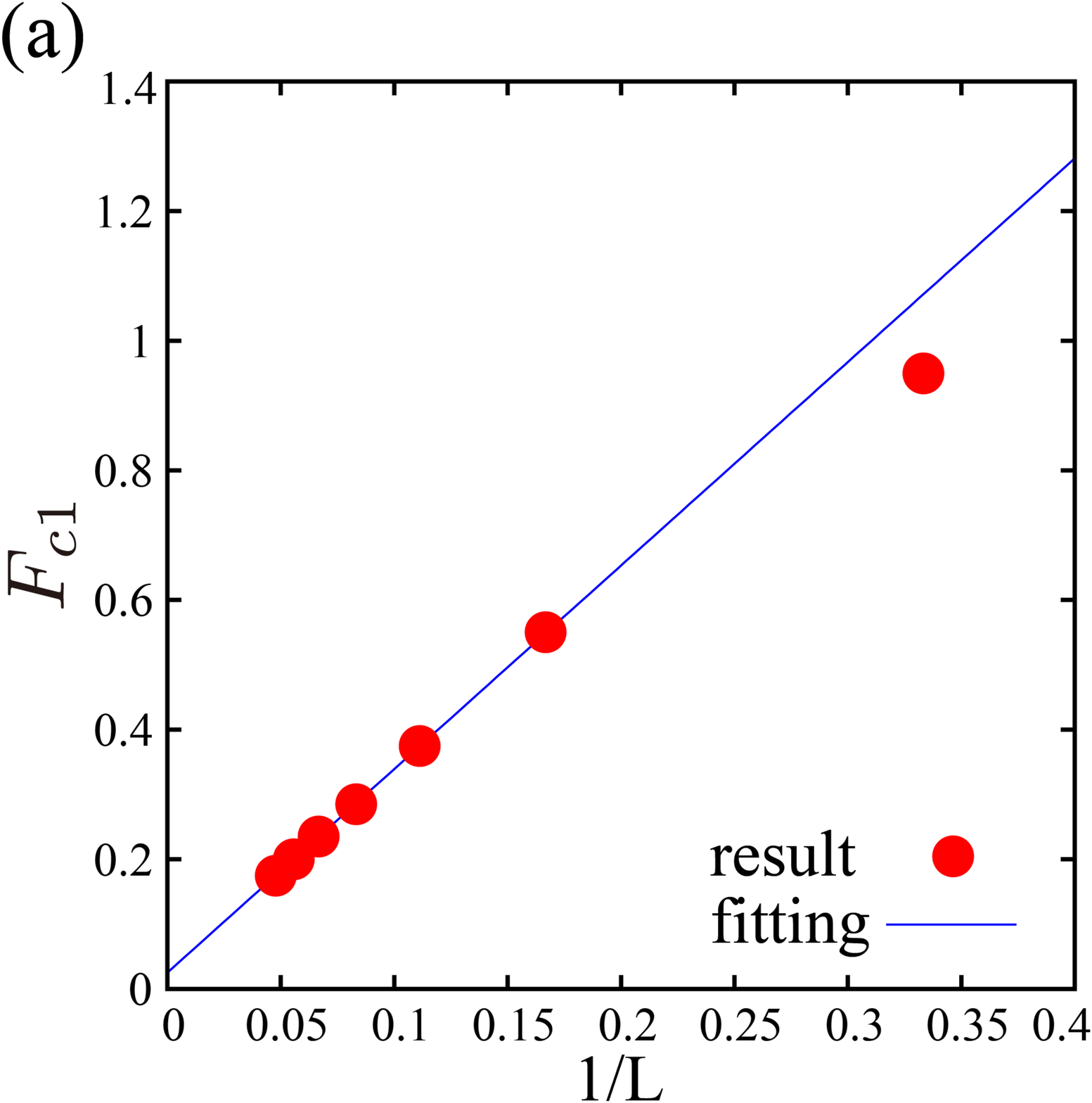}
\includegraphics[width=4.2cm]{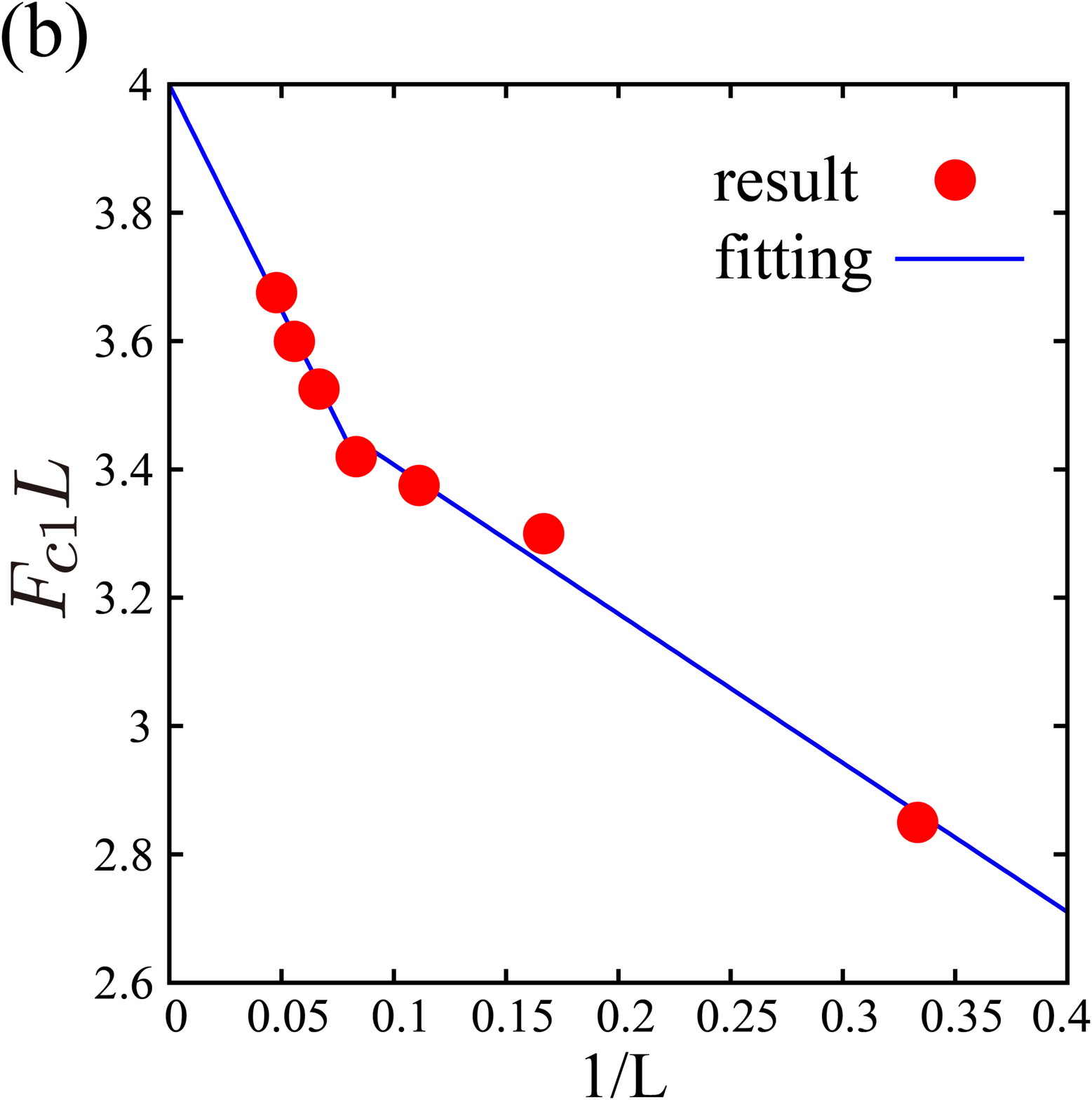}
\end{center}
\caption{Finite-size scaling of the critical gradient of topological state, 
(a) $F_{c1}$ and (b) $F_{c1}L$. 
}
\label{systemsize2}
\end{figure}

In this Appendix, we study the system-size dependence of the Chern number.
We first show the calculations in $L=15$ and $21$ in Fig.~\ref{systemsize1}.
Compared with the calculations in Fig.~\ref{CHN1}, we find that the Chern number has only
small dependence on the system size, whereas the energy gap exhibits a slightly different
behavior in the systems $L=15, 18$ and $L=21$ as a function of $F$.
The `critical regime' is slightly smaller in the $L=15$ system, $0.2<F<0.42$,
than in the $L=18$ system.
This result implies that there exists a finite regimes between the $F_{c1}$ and $F_{c2}$
for the limit $L\to \infty$.

We calculated the Chern number in smaller system sizes to extrapolate the critical value
$F_{c1}$ for  $L\to \infty$.
The result is shown in Fig.~\ref{systemsize2}.
Fig.~\ref{systemsize2} (a) shows that $F_{c1}$ changes its behavior at $L=6$,
and the exprapolation by $L\ge 6$ give an estimation such as $F_{c1} \to 0$
for $L\to \infty$.
This result is not so surprising because in the limit $L\to \infty$,
the difference in the gradient potential between two edges tends to $\infty$
for $F\neq 0$.
Fig.~\ref{systemsize2} (b) shows $(F_{c1}L)$ vs. ${1/L}$.
The result indicates that the topological state survives in the limit $L\to \infty$
as long as the difference in the gradient potential between two edges is finite.

\section{System size dependence for Phase diagram of 
the topological phase in Fig.~\ref{WFphase1}}

\setcounter{figure}{0}

\begin{figure}[h]
\centering
\begin{center}
\vspace{0.5cm}
\includegraphics[width=5.5cm]{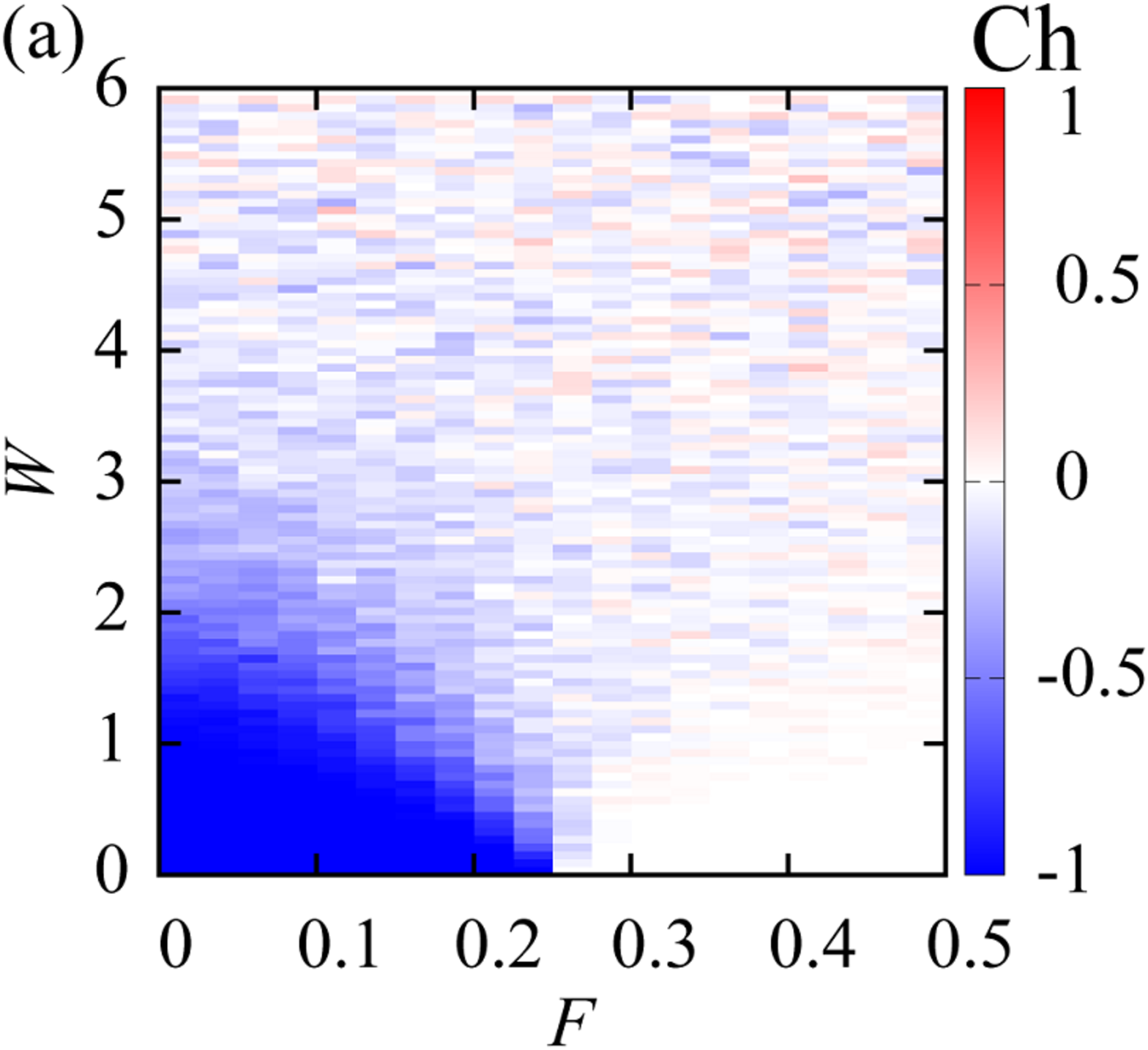}
\includegraphics[width=5.5cm]{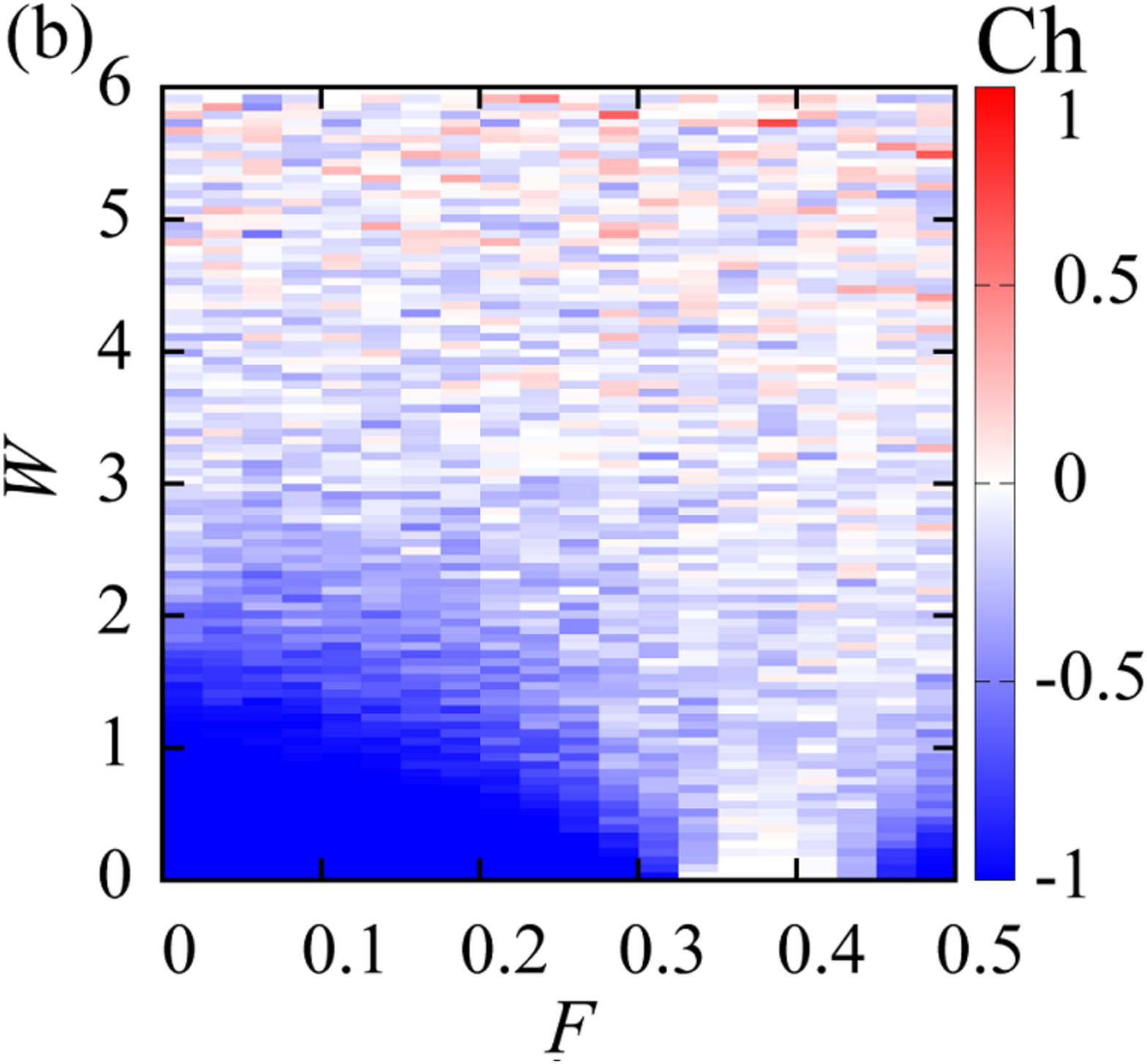}
\end{center}
\vspace{-0.5cm}
\caption{Phase diagram of the topological phase obtained for system size $L=15$,
(a) for the linear-gradient potential, (b) for the $V$-shape potential.
Phase diagrams for $L=18$ are shown in Fig.~\ref{WFphase1}, and only small system-size
depenedence is observed. 
}
\label{systemsize3}
\end{figure}

As in appendix A, we show the system-size dependence of the ground-state phase diagram
with the topological order obtained by calculating Chern number in the system $W>0$.
Figure~\ref{systemsize3} displays the topological phase in the $(F-W)$ plain 
obtained for $L=15$ system under the linear and $V$-shape potentials,
which should be compared with Fig.~\ref{WFphase1}. 
For the case of the linear potential, two phase diagrams of $L=18$ and $L=15$ are
almost the same.
On the other hand in the $V$-shape potential case, the location of the finite 
Chern number regime in $W\ll 1$ and $F>0.3$ slightly changes. 
Anyway, we conclude that the phase diagram of the topological state has only small
system-size dependence.

\newpage



\begin{thebibliography}{99}
\bibitem{ED5}
Here, it is noted that our prepared Hilbert space does not conserve the total $z$-component spin 
$\sum_{i}S^{z}_{i}$.
\bibitem{ED6}
In what follows, the Hilbert space in the ED calculation is restricted to the sector with
the total $z$-component spin fixed to $m_zL$.
\bibitem{FN1}
We should remark that the definition of the parameter $F$ in Ref.~\cite{Refael}
is different from ours by a factor of 2, although this is not explicitly announced in \cite{Refael}.
See also recent paper, D. S. Bhakuni and A. Sharma, arXiv: 1909.10542.
\end{thebibliography}
\end{document}